\documentclass{aa}
\usepackage[colorlinks=true,linkcolor=blue,citecolor=blue,urlcolor=blue]{hyperref}
\usepackage{graphicx}
\usepackage{txfonts}
\usepackage{lipsum}
\usepackage{subcaption}        
\usepackage{lscape}            
\usepackage{placeins}
\usepackage{wrapfig}
\usepackage[dvipsnames]{xcolor}

\begin{document}

\title{Chemical structure of the solar neighbourhood with DESI DR1}

\author{Yi-De Jing\inst{1,2}\corrauth{jingyide@qhu.edu.cn}
        \and Duojie Jia\inst{3,4}\corrauth{jiadj@nwnu.edu.cn} 
        \and Yulong Gao\inst{1,2}
        \and Zheng-Jun Du\inst{5,6}
        \and Zenghui Zou\inst{1,2}
        \and Liyuan Lu\inst{1,2}}

\institute{Department of Physics and Astronomy, Qinghai University, Xining, 810016, China
\and Research Center of Astronomy, Qinghai University, Xining, 810016, China
\and General Education Center, Qinghai Institute of Technology, No. 2, Xining, 810016, China
\and Lanzhou Center for Theoretical Physics, Key Laboratory of Theoretical Physics of Gansu Province, Key Laboratory of Quantum Theory and Applications of MoE, Gansu Provincial Research Center for Basic Disciplines of Quantum Physics, Lanzhou University, Lanzhou 730000, China
\and School of Computer Technology and Application, Qinghai University, Xining 810016, China
\and Qinghai Provincial Key Laboratory of Media Integration Technology and Communication, Qinghai University, Xining 810016, China}

\date{Received August 8, 2026}
 
  \abstract
{Stellar chemical abundances encode the formation and evolutionary history of the Galactic disc. The bimodality observed in the $[\alpha/\mathrm{Fe}]$--$[\mathrm{Fe}/\mathrm{H}]$ plane has been widely interpreted as evidence of distinct thin- and thick-disc populations.}
{We analysed the chemical, spatial, and kinematic properties of stars in the solar neighbourhood using the Dark Energy Spectroscopic Instrument (DESI) Data Release 1 combined with \textit{Gaia} astrometry, aiming to characterize the low-$\alpha$ and high-$\alpha$ populations and their evolutionary connections.}
{We constructed a high-quality sample of FGK dwarf stars by applying rigorous selection criteria to stellar parameters, abundances, and distances. We employed Gaussian mixture modelling to statistically separate low-$\alpha$ and high-$\alpha$ stars and examined their metallicity distribution functions (MDFs), vertical and radial gradients, and kinematics.}
{The DESI solar neighbourhood sample reveals a clear bimodality. The low-$\alpha$ stars dominate near the Galactic mid-plane, with a broad MDF peaking at $[\mathrm{Fe/H}] \sim -0.2$ and negative vertical and radial metallicity gradients of $\mathrm{d[Fe/H]}/\mathrm{d}|Z| \sim -0.06$ dex kpc$^{-1}$ and $\mathrm{d[Fe/H]}/\mathrm{d}R \sim -0.033$ dex kpc$^{-1}$, respectively. In contrast, the high-$\alpha$ population exhibits a narrower MDF centred at $[\mathrm{Fe}/\mathrm{H}] \sim -0.5$, which has a steeper vertical metallicity gradient ($\mathrm{d[Fe/H]}/\mathrm{d}|Z| \sim -0.10$ dex kpc$^{-1}$), an approximately flat radial profile of metallicity, and dynamically hotter kinematics. Nevertheless, the Toomre diagram shows substantial overlap between these chemically defined populations in velocity space, indicating that chemical abundance does not uniquely map to distinct kinematic components.}
{These results indicate that the low-$\alpha$ and high-$\alpha$ populations underwent distinct chemical enrichment histories despite their substantial overlap in present-day kinematics. The observed chemo-dynamical properties point to the combined effects of chemical enrichment, radial redistribution, and secular dynamical evolution, underscoring the necessity of integrating stellar chemistry with kinematics to constrain the evolution of the Galactic disc.}


\maketitle

\nolinenumbers 

\section{Introduction}

Understanding the formation and evolution of the Milky Way disc is a central goal of modern Galactic archaeology. A key open question is whether the observed chemical structure reflects discrete formation episodes or a continuous evolutionary process. Stellar elemental abundances provide a fossil record of the Galaxy’s star formation history and chemical enrichment \citep{Freeman2002, RixBovy2013}.

Of the various chemical diagnostics, the $\alpha$-element abundance ratio relative to iron, $[\alpha/\mathrm{Fe}]$, serves as a critical probe. $\alpha$-elements are predominantly synthesized in core-collapse supernovae on short timescales, whereas iron is primarily contributed by Type Ia supernovae on delayed timescales. Consequently, $[\alpha/\mathrm{Fe}]$ encodes the timescales of star formation and chemical enrichment, making it a powerful diagnostic for the early evolution of the Galactic disc \citep{Tinsley1979, MatteucciGreggio1986, Nomoto2013}.

 One of the most striking observational results from Galactic chemical evolution studies is the bimodal distribution of disc stars in the $[\alpha/\mathrm{Fe}]$--$[\mathrm{Fe}/\mathrm{H}]$ plane. Early high-resolution spectroscopic studies of nearby stars revealed two distinct chemical sequences: a metal-poor, $\alpha$-enhanced population and a metal-rich, low-$\alpha$ population \citep{Fuhrmann1998, Gratton2000, Bensby2003, Reddy2006}. These sequences were traditionally associated with the thick and thin discs, respectively, suggesting that the Milky Way disc underwent at least two phases of star formation characterized by different enrichment timescales and dynamical conditions \citep{Gilmore1983}.

Large-scale spectroscopic surveys over the past decade have firmly established the ubiquity of this chemical structure throughout the Galaxy. Results from the Apache Point Observatory Galactic Evolution Experiment (APOGEE), \textit{Gaia}-ESO Survey, and the Galactic Archaeology with HERMES (GALAH) survey show that the high-$\alpha$ sequence is relatively narrow in chemical space and occupies a limited metallicity range, while the low-$\alpha$ sequence spans a broader range in $[\mathrm{Fe}/\mathrm{H}]$ and dominates the stellar population near the solar radius \citep{Nidever2014, Hayden2015, Haywood2013, Buder2019, Weinberg2019}. Moreover, the relative prominence of the two sequences varies systematically with Galactocentric radius and height above the mid-plane, pointing to a complex interplay between star formation history, disc heating, and radial migration \citep{Queiroz2020, Anders2023, Queiroz2023}.

Despite this observational progress, it remains unclear whether the observed bimodality reflects truly distinct stellar populations or instead arises from a continuous distribution shaped by time-dependent star formation and radial migration. This ambiguity lies at the heart of current debates on the formation of the Galactic disc. Recent studies have emphasized that the low-$\alpha$ and high-$\alpha$ populations are not fully disjoint in either spatial or kinematic space. Instead, they exhibit substantial overlap, particularly in the solar neighbourhood, challenging the notion of a strict thin--thick disc dichotomy \citep{Bovy2012, Mackereth2019, Trick2019}. In this view, the Galactic disc may represent a continuous structure in chemo-dynamical space, with chemically defined populations reflecting differences in star formation timescales rather than discrete formation events \citep{Palla2020, Sharma2021}.

A wide range of theoretical scenarios have been proposed to explain the origin of the observed chemical structure. These include an early phase of rapid star formation followed by a decline in star formation efficiency \citep{Chiappini1997, Haywood2016}, gas-rich mergers at high redshifts that dynamically heat the early disc \citep{Brook2004, Quinn1993}, and inside-out disc growth combined with radial migration driven by spiral arms and the Galactic bar \citep{Schonrich2009, Minchev2013, Frankel2018}. More recent models invoke time-dependent star formation efficiencies or multiple gas accretion episodes to reproduce the detailed morphology of the $[\alpha/\mathrm{Fe}]$--$[\mathrm{Fe}/\mathrm{H}]$ plane \citep{Spitoni2019, Johnson2021, Khanna2022}.

The solar neighbourhood provides a particularly valuable laboratory for such studies. Its proximity allows for high-precision astrometry from the \textit{Gaia} mission, delivering accurate distances and proper motions that enable detailed investigations of the local disc’s spatial, kinematic, and vertical structure \citep{Gaia2016, Gaia2018, Gaia2021, Gaia2023}. As a result, the local disc has long served as a benchmark for studies of disc kinematics, chemical evolution, and stellar populations \citep{Edvardsson1993, Adibekyan2012, DelgadoMena2019}.

In recent years, increasing attention has been paid to vertical metallicity gradients and their dependence on chemical abundance patterns. Vertical gradients provide sensitive diagnostics of disc heating, secular evolution, and radial migration, and how they differ between low-$\alpha$ and high-$\alpha$ populations offers important constraints on disc formation models \citep{Minchev2014, Martig2016, Sharma2021, Queiroz2023}. Whether these populations exhibit significantly different gradients, or instead reflect a smooth and continuous disc structure, remains an open question.

The Dark Energy Spectroscopic Instrument (DESI) offers a new and independent opportunity to address these issues. Although primarily designed for cosmological surveys, DESI has obtained optical spectra for millions of Galactic stars with homogeneous data reduction and stellar parameter determination. DESI Data Release 1 (DR1) provides atmospheric parameters and chemical abundances through the Milky Way Survey stellar value-added catalogue, based on the IRON spectroscopic production and dedicated stellar-analysis pipelines, enabling large-scale chemo-dynamical studies of the Milky Way disc \citep{DESI2023a}. As an optical survey, DESI complements infrared surveys such as APOGEE and offers an independent assessment of the Galaxy’s chemical structure, helping to identify and quantify potential systematic effects in different datasets \citep{Jofre2019}.

In this work, we exploited DESI DR1 stellar data combined with \textit{Gaia} astrometry to investigate the chemical, spatial, and kinematic properties of stars in the solar neighbourhood. We focused on characterizing the chemically defined low-$\alpha$ and high-$\alpha$ populations and exploring the relationship between their chemical abundances, spatial distributions, and kinematic properties. Specifically, we address three main questions: (i) how the low-$\alpha$ and high-$\alpha$ populations are distributed in the $[\alpha/\mathrm{Fe}]$--$[\mathrm{Fe}/\mathrm{H}]$ plane and how their metallicity distribution functions (MDFs) differ; (ii) how their vertical and radial metallicity structures compare; and (iii) how their kinematic properties relate to their chemical classifications.

This paper is organized as follows. In Sect.~\ref{SEc_Data_and_sample_selection} we describe the DESI and \textit{Gaia} datasets and the construction of the stellar sample. The main observational results are presented in Sect.~\ref{SEC_Results}. The implications of our results for the formation and evolution of the Galactic disc are discussed in Sect.~\ref{SEC_Discussion}, where the main conclusions of this work are also summarized.

\section{Data and sample selection}\label{SEc_Data_and_sample_selection}

\subsection{DESI and Gaia data}

This study is based on a combination of spectroscopic data from DESI \citep{DESI2024} and astrometric data from the \textit{Gaia} mission \citep{Gaia2016, Gaia2023}. DESI is a wide-field, multi-object spectroscopic survey designed to obtain optical spectra for tens of millions of targets over a large fraction of the sky. Within DESI, the Milky Way Survey provides a homogeneous and high-quality spectroscopic dataset for Galactic stars, including stellar radial velocities, atmospheric parameters, and chemical abundances.

The DESI stellar parameter pipeline delivers estimates of effective temperature ($T_{\mathrm{eff}}$), surface gravity ($\log g$), and chemical abundances, including iron abundance $[\mathrm{Fe}/\mathrm{H}]$ and $\alpha$-element enhancement $[\alpha/\mathrm{Fe}]$. These parameters are derived from DESI optical spectra covering the wavelength range 3600--9800\,\AA. The DESI spectrograph employs three channels spanning the blue, red, and near-infrared wavelength regions, with spectral resolutions ranging from approximately $R\sim2000$ to $R\sim5500$ across the full wavelength coverage \citep{DESI2024}. The broad wavelength coverage includes numerous absorption features sensitive to stellar atmospheric parameters and chemical abundances, enabling reliable determinations of $[\mathrm{Fe}/\mathrm{H}]$ and $[\alpha/\mathrm{Fe}]$ for FGK-type stars.

Astrometric information is taken from the DESI value-added catalogues, which incorporate \textit{Gaia} Early Data Release 3 measurements. Distances and proper motions are therefore based on \textit{Gaia} astrometry, but are adopted directly from the DESI pipeline products without further modification. The high precision of \textit{Gaia} data ensures reliable spatial information for stars in the solar neighbourhood, making the combined DESI spectroscopic and \textit{Gaia}-based astrometric dataset well suited for studies of Galactic chemical structure.

\subsection{Cross-match and initial quality selection}

We used data from the DESI DR1 value-added catalogues \citep{DESI2024}, which combine spectroscopic and astrometric information for Milky Way stars. Our analysis is based on two complementary datasets.

The first dataset is the IRON catalogue (\texttt{iron-yr1-v2.1}), from which we adopted astrometric and distance information, including right ascension, declination, proper motions, heliocentric distances (DIST), and the re-normalised unit weight error (RUWE), the astrometric quality indicator. These quantities are based on \textit{Gaia} astrometry within the DESI pipeline.

The second dataset is the Milky Way Survey all-sky catalogue (\texttt{mwsall-pix-iron}), which provides stellar atmospheric parameters and chemical abundances derived from DESI spectra \citep{DESIPipeline2023}. In particular, we used metallicity $[\mathrm{Fe}/\mathrm{H}]$, $\alpha$-element abundance $[\alpha/\mathrm{Fe}]$, effective temperature ($T_{\mathrm{eff}}$), surface gravity ($\log g$), radial velocity, and their associated uncertainties.

The two catalogues were matched using the unique DESI target identifier (\texttt{TARGETID}), ensuring a one-to-one correspondence between astrometric and spectroscopic information. Only stars with valid entries in both catalogues were retained in the analysis.

To ensure reliable measurements, we applied several quality cuts. On the spectroscopic side, we defined a combined signal-to-noise ratio as
\begin{equation}
        \mathrm{S/N} = \sqrt{\mathrm{SN}_B^2 + \mathrm{SN}_R^2 + \mathrm{SN}_Z^2},
\end{equation}
and required $\mathrm{S/N} \ge 10$.
We further required
$\sigma_{\mathrm{[Fe/H]}} < 0.2$ dex and
$\sigma_{[\alpha/\mathrm{Fe}]} < 0.1$ dex,
and restricted the abundance range to
$[\mathrm{Fe}/\mathrm{H}] > -2.5$ and
$-0.2 < [\alpha/\mathrm{Fe}] < 0.6$.

On the astrometric side, we removed sources with potentially problematic solutions by requiring $\mathrm{RUWE} < 1.4$. We also retained only stars with positive distance estimates ($\mathrm{DIST} > 0$).

The sample-selection criteria adopted in this work are summarized in Table~\ref{tab:sample_selection}. After applying these quality cuts, the initial matched sample contains approximately $2.6 \times 10^{6}$ stars with reliable spectroscopic and astrometric measurements. Subsequent FGK dwarf and solar neighbourhood selections yield a final sample of 334\,255 stars for the analysis presented in this paper.

\begin{table}[ht!]
        \centering
        \caption{Sample-selection criteria adopted in this work.}
        \begin{tabular}{ll}
                \hline
                \textbf{Selection category} & Adopted criterion \\
                \hline
                Spectroscopic quality       & $\sqrt{SN_B^2 + SN_R^2 + SN_Z^2} \ge 10$ \\
                Abundance uncertainty        & $\sigma_{[\mathrm{Fe/H}]} < 0.2$ dex \\
                Abundance uncertainty        & $\sigma_{[\alpha/\mathrm{Fe}]} < 0.1$ dex \\
                Metallicity range            & $[\mathrm{Fe/H}] > -2.5$ \\
                Alpha-abundance range        & $-0.2 < [\alpha/\mathrm{Fe}] < 0.6$ \\
                Astrometric quality         & RUWE $< 1.4$ \\
                Distance selection           & DIST $> 0$ \\
                Effective temperature        & $4500 < T_{\rm eff} < 6500$ K \\
                Surface gravity              & $\log g > 3.5$ \\
                Solar neighbourhood selection & $D_\odot \le 1.5$ kpc \\
                Vertical-height selection    & $|Z| \le 0.8$ kpc \\
                \hline
                Initial sample size           & 2\,573\,659 \\
                Final sample size             & 334\,255 \\
                \hline
        \end{tabular}
        \label{tab:sample_selection}
\end{table}

\subsection{Transformation to Galactocentric coordinates}

For spatial selection and visualization, stellar positions and velocities were transformed from equatorial coordinates into a right-handed Galactocentric Cartesian coordinate system using the \texttt{Astropy} framework \citep{Astropy2018}. The transformation was performed using sky positions, heliocentric distances, proper motions, and radial velocities. The heliocentric distances were taken directly from the DESI IRON catalogue.

We adopted a solar position of $R_\odot = 8.2$~kpc \citep{Gravity2019} and $Z_\odot = 0.02$~kpc \citep{BennettBovy2019} above the Galactic mid-plane. These values are consistent with recent measurements and have negligible impact on our results given the limited spatial extent of the solar neighbourhood sample. The Galactocentric coordinates $(X, Y, Z)$ are defined such that the Galactic centre is located at $(0, 0, 0)$, the $X$-axis points from the Galactic centre towards the Sun, and the $Z$-axis points towards the north Galactic pole.

\subsection{Solar neighbourhood selection}

To focus on the local Galactic environment and to minimize the influence of large-scale radial and vertical gradients, we restricted our analysis to stars located in the solar neighbourhood \citep{BlandHawthorn2016, Gaia2023}. The heliocentric distance ($D_\odot$) was computed from the Galactocentric Cartesian coordinates of each star using the adopted solar position. This ensures a consistent spatial definition of the solar neighbourhood within the Galactocentric reference frame.

We selected stars within
\begin{equation}
        D_\odot \le 1.5~\mathrm{kpc},
\end{equation}
which provides a balance between maintaining a sufficiently large sample size and limiting the effects of large-scale radial gradients. Within this distance range, the sample remains relatively local while still containing enough stars for statistically robust analysis. In addition, we required
\begin{equation}
        |Z| \le 0.8~\mathrm{kpc}
\end{equation}
to restrict the sample to the Galactic disc and to exclude stars at large vertical distances from the mid-plane. This criterion reduces the influence of halo contamination and ensures that the selected stars primarily trace the disc population.

This combined spatial selection results in a sample that is both spatially localized and representative of the solar neighbourhood disc. For analyses focusing on radial trends, we adopted this vertical cut ($|Z| \le 0.8$~kpc) to ensure a well-defined local disc sample. However, when investigating vertical metallicity gradients and constructing the $|Z|$ distribution functions, we extended the sample to larger vertical distances (up to $|Z| \sim 1.5$~kpc) in order to better capture the full vertical structure of the disc and improve the statistical robustness of the derived gradients. We verified that restricting the analysis to $|Z| \le 0.8$~kpc does not qualitatively change the inferred trends, although it reduces the dynamic range in $|Z|$. Therefore, the use of an extended vertical range does not bias our main conclusions.

\begin{figure}[ht!]
        \centering
        \vspace{-0.2cm}
        \includegraphics[width=0.9\hsize]{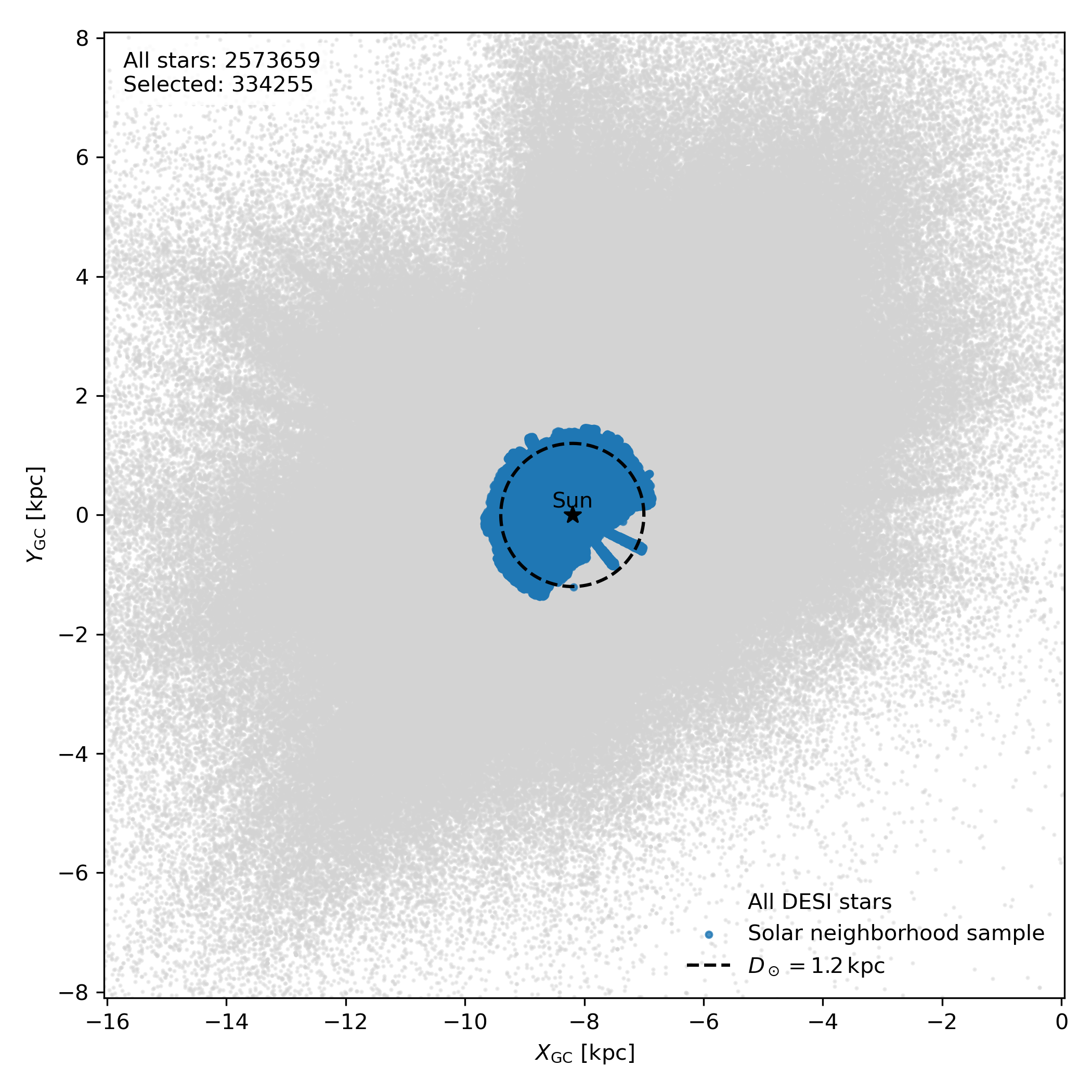}
        \caption{Spatial distribution of DESI stars in the Galactocentric $X$--$Y$ plane. Grey points show all DESI stars, while blue points indicate the selected solar neighbourhood sample within $D_\odot < 1.5\,\mathrm{kpc}$. The dashed black circle marks the selection boundary, and the Sun is indicated by a black star.}
        \label{Fig:sample_selection_XY}
\end{figure}

Figure~\ref{Fig:sample_selection_XY} illustrates the spatial distribution of all DESI stars in the Galactocentric $X$--$Y$ plane and highlights the selected solar neighbourhood sample. The non-uniform spatial distribution reflects the DESI survey footprint and target selection strategy rather than any intrinsic asymmetry in the solar neighbourhood.

Because our analysis focuses on relative chemical trends rather than absolute number densities, and the low-$\alpha$ and high-$\alpha$ populations are subject to the same selection effects, this non-uniformity is not expected to bias our main conclusions. In particular, comparisons between chemically defined populations remain robust against such spatial incompleteness. The solar neighbourhood selection yields approximately $3.3 \times 10^{5}$ stars, providing a large and statistically significant sample for subsequent analysis.

\subsection{Stellar type selection: FGK stars}

To ensure accurate and homogeneous stellar parameter estimates, we restricted our analysis to FGK-type stars. These stars have well-understood atmospheres and exhibit numerous metal absorption lines in optical spectra, making them reliable tracers of Galactic chemical evolution \citep{BlandHawthorn2016}.

The FGK sample was selected using effective temperature and surface gravity cuts of $4500~\mathrm{K} < T_{\mathrm{eff}} < 6500~\mathrm{K}$ and $\log g > 3.5$. These criteria exclude hot early-type stars and evolved giants, yielding a sample dominated by main-sequence and turn-off stars. Main-sequence and turn-off stars generally provide more reliable atmospheric parameters and chemical abundances due to their simpler atmospheric structures and smaller systematic uncertainties. In addition, their distances and kinematics are typically better constrained, making them particularly suitable for studies of the chemo-dynamical structure of the Galactic disc. Similar selection criteria have been widely adopted in previous studies of Galactic disc chemo-dynamical evolution \citep{Haywood2013,Feuillet2019,Xiang2022}.

These parameter ranges are also well matched to the regime in which the DESI stellar parameter pipeline delivers the most robust atmospheric parameters and chemical abundances. We verified that modest variations in these cuts do not qualitatively affect our main results.

\subsection{Metallicity definition and uncertainties}

We adopted the stellar metallicity $[\mathrm{Fe}/\mathrm{H}]$ provided by the DESI stellar parameter pipeline, derived from template fitting to medium-resolution optical spectra. These measurements were calibrated using reference stellar samples and are expected to be reliable within the FGK parameter regime.

To ensure high-quality metallicity estimates, we required $\sigma_{\mathrm{[Fe/H]}} < 0.2$~dex and excluded stars with poorly constrained measurements. A similar quality cut was applied to the $\alpha$-element abundance uncertainties. These criteria minimize the impact of measurement noise and ensure that the observed metallicity dispersion primarily reflects intrinsic astrophysical variations.

We confirmed that adopting slightly different uncertainty thresholds does not significantly change the overall metallicity trends or gradients. Although systematic uncertainties in spectroscopic abundance determinations may still be present, they are expected to affect the low-$\alpha$ and high-$\alpha$ populations in a similar way. Therefore, relative comparisons between the two populations remain robust.

\subsection{Gaussian mixture model analysis}

To characterize the structure of the chemical abundance distribution in the $[\alpha/\mathrm{Fe}]$--$[\mathrm{Fe}/\mathrm{H}]$ plane, we applied a Gaussian mixture model (GMM) to the solar neighbourhood sample. The abundance distribution of disc stars is known to deviate significantly from a single Gaussian and typically exhibits multiple
sequences and extended features, reflecting the presence of stellar populations with different formation and chemical enrichment histories \citep{Hayden2015, Frankel2018}.

In this work, we adopted a two-component GMM and fitted it directly to the two-dimensional abundance space defined by $[\mathrm{Fe}/\mathrm{H}]$ and $[\alpha/\mathrm{Fe}]$. The model parameters were estimated using the expectation--maximization algorithm as implemented in the \texttt{scikit-learn} package \citep{Pedregosa2011}. The default configuration was adopted, which employs full covariance matrices for each component and therefore enables correlations between $[\mathrm{Fe}/\mathrm{H}]$ and $[\alpha/\mathrm{Fe}]$ to be determined.

Each star was assigned to one of the two Gaussian components based on the maximum-likelihood classification returned by the model. Compared to simple threshold-based cuts in chemical space, this approach provides a more objective way to separate the two dominant sequences in the abundance plane.

To identify the physical meaning of the two components, we compared their mean $[\alpha/\mathrm{Fe}]$ values and defined the component with the higher mean $[\alpha/\mathrm{Fe}]$ as the high-$\alpha$ population; the other component was classified as the low-$\alpha$ population. This procedure yields a robust separation of the two dominant sequences commonly associated with the chemically defined thick and thin disc components.

We emphasize that the number of Gaussian components is fixed to two, motivated by the clear bimodal structure visible in the abundance distribution. The GMM is therefore used as a descriptive statistical tool to quantify this structure and to define chemically distinct subsamples, rather than to identify strictly discrete stellar populations. In particular, stars located in the overlap region between the two components may be assigned differently depending on small variations in the model, reflecting the continuous nature of chemical evolution and the effects of radial migration in the Galactic disc \citep{Hayden2015, Frankel2018}.

We verified that the resulting separation remains stable under reasonable variations in the adopted sample-selection criteria. Specifically, repeating the analysis with different cuts in surface gravity, effective temperature, and signal-to-noise ratio yields nearly identical centroid positions and population fractions for the low-$\alpha$ and high-$\alpha$ components. The maximum centroid shift is smaller than 0.013 dex in all tested cases (see Appendix~\ref{Appendix_A1}), demonstrating that the observed chemical bimodality is robust against the adopted sample-selection criteria. The robustness of the GMM decomposition, including the effects of model initialization, component number, and sampling fluctuations, is further examined in Appendix~\ref{Appendix}.

\section{Results}\label{SEC_Results}

\subsection{Chemical abundance distribution in the solar neighbourhood}

\begin{figure*}[ht!]
        \centering
        \vspace{-0.2cm}
        \resizebox{17cm}{6cm}
        {\includegraphics {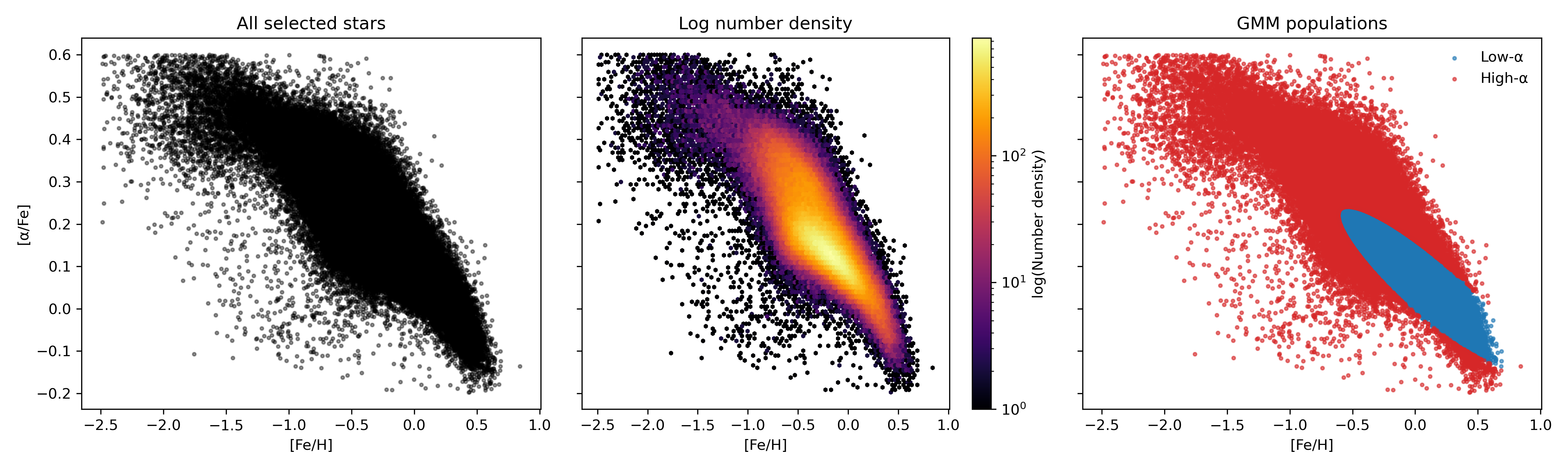}}
    \caption{Distribution of the DESI solar neighbourhood sample in the $[\alpha/\mathrm{Fe}]$--$[\mathrm{Fe}/\mathrm{H}]$ plane. \textit{Left}: Raw distribution of all selected stars, shown as individual data points. \textit{Middle}: Two-dimensional number density map of the same sample, with colours representing the $\log_{10}$ of the stellar number density. \textit{Right}: Chemical populations identified using a GMM. Stars assigned to the low-$\alpha$ and high-$\alpha$ components are displayed in different colours.}
        \label{Fig:Fig_summary_alpha_feh.png}
\end{figure*}

Figure~\ref{Fig:Fig_summary_alpha_feh.png} provides an overview of the chemical abundance properties of the DESI solar neighbourhood sample in the $[\alpha/\mathrm{Fe}]$--$[\mathrm{Fe}/\mathrm{H}]$ plane. The three panels illustrate complementary views of the same dataset, highlighting the progression from the raw stellar distribution to a statistically motivated chemical classification.

The left panel shows the distribution of individual stars after applying the quality cuts, stellar parameter selections, and spatial constraints described in Sect.~\ref{SEc_Data_and_sample_selection}. Although individual measurements exhibit a non-negligible scatter, the distribution is clearly structured and departs from a smooth, single-component population \citep{Fuhrmann1998, Bensby2003, Nidever2014}. The majority of stars occupy a sequence at low $[\alpha/\mathrm{Fe}]$, typically $[\alpha/\mathrm{Fe}] \lesssim 0.1$, spanning a broad metallicity range from $[\mathrm{Fe}/\mathrm{H}] \approx -0.8$ to super-solar values. In addition, a clear enhancement in stellar density is visible at elevated $[\alpha/\mathrm{Fe}]$ values ($[\alpha/\mathrm{Fe}] \gtrsim 0.15$) and subsolar metallicities, with a concentration around $[\mathrm{Fe}/\mathrm{H}] \sim -0.4$ \citep{Adibekyan2012, Haywood2013}. This overdense structure indicates the presence of a chemically distinct population that cannot be explained as a simple extension of the low-$\alpha$ sequence, and provides clear evidence of at least two chemically distinct stellar populations in the solar neighbourhood.

To better visualize the intrinsic structure of the distribution, the middle panel of Fig.~\ref{Fig:Fig_summary_alpha_feh.png} shows the stellar number density in the $[\alpha/\mathrm{Fe}]$--$[\mathrm{Fe}/\mathrm{H}]$ plane, colour-coded by $\log_{10}$ of the number density. In this representation, the bimodal chemical structure of the disc becomes unambiguous \citep{Nidever2014, Hayden2015}. This bimodality cannot be explained by observational scatter alone, indicating an intrinsic chemical dichotomy in the Galactic disc. The low-$\alpha$ sequence dominates the solar neighbourhood and exhibits its highest density near $[\mathrm{Fe}/\mathrm{H}] \sim 0.0$ and $[\alpha/\mathrm{Fe}] \sim 0.0$, with a relatively broad metallicity spread. In contrast, the high-$\alpha$ sequence forms a narrower ridge, centred at $[\alpha/\mathrm{Fe}] \sim 0.2$--0.3, and is confined to a more limited metallicity range, peaking at $[\mathrm{Fe}/\mathrm{H}] \sim -0.4$. The two sequences overlap over a finite metallicity interval, approximately $-0.6 \lesssim [\mathrm{Fe}/\mathrm{H}] \lesssim -0.2$, where stars with similar metallicities exhibit significantly different $[\alpha/\mathrm{Fe}]$ abundances \citep{Bovy2012, Mackereth2019}. This overlap demonstrates that a separation based solely on metallicity would lead to substantial cross-contamination between the two populations. This reflects the complex chemical enrichment history of the Galactic disc. This motivates a quantitative separation of the two sequences using a GMM.

The right panel presents the result of a GMM analysis applied to the two-dimensional abundance space. The GMM identifies two statistically distinct components whose centroids and dispersions closely follow the ridge lines of the low-$\alpha$ and high-$\alpha$ sequences revealed by the density map. Stars are coloured according to the component label assigned by the GMM, illustrating that the separation follows the global structure of the abundance distribution rather than local fluctuations or measurement noise. The relative normalization of the two components indicates that the low-$\alpha$ population accounts for approximately 59.3$\%$ of the solar neighbourhood sample, whereas the high-$\alpha$ population contributes about 40.7$\%$ (see Appendix~\ref{Appendix_A1} and Table~\ref{Tab_gmm_robustness}). We note that the GMM provides a statistical approximation to the underlying distribution and does not imply that the two components represent strictly discrete stellar populations.

The robustness of the GMM decomposition is examined in Appendix~\ref{Appendix}, where we explore the impact of different initializations, test the behaviour of the model under alternative component numbers using the Bayesian information criterion (BIC), and assess the stability of the results through bootstrap resampling. These tests confirm that the two-component model provides a stable and statistically justified description of the abundance distribution.

This GMM-based classification provides a reproducible framework for defining chemically distinct disc populations, which are used in the subsequent analysis. The bimodal structure observed in the DESI sample is consistent with that reported in previous large spectroscopic surveys such as APOGEE and GALAH \citep{Nidever2014, Hayden2015, Buder2019}, confirming that the chemical dichotomy of the Galactic disc is robust across different datasets.

\subsection{Metallicity distribution functions}

\begin{figure}[ht!]
        \centering
        \vspace{-0.2cm}
        \includegraphics[width=\hsize]{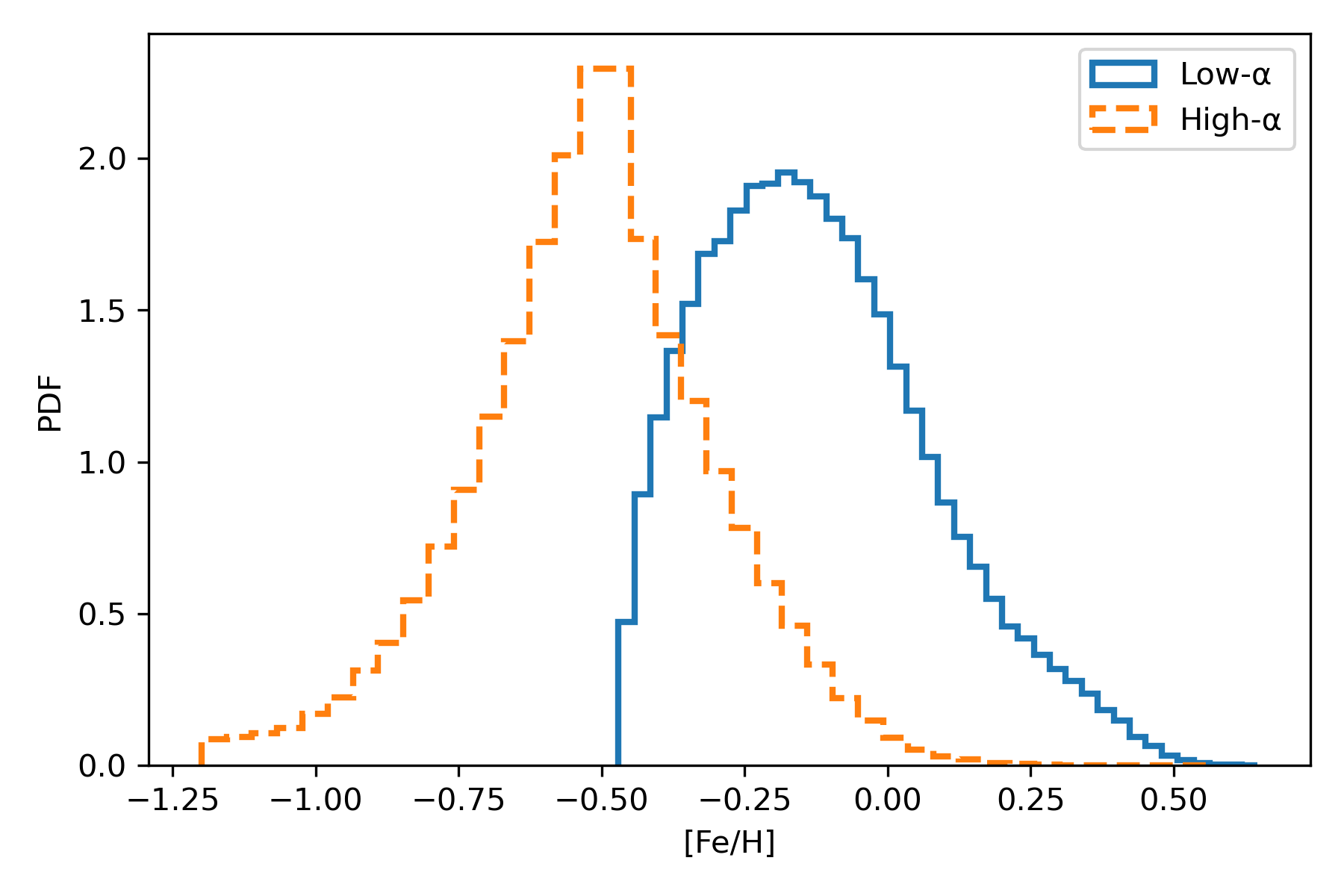}
    \caption{MDFs of the chemically defined low-$\alpha$ (blue) and high-$\alpha$ (red) populations in the solar neighbourhood. The distributions are normalized to unit area.}
        \label{Fig:fig2_mdf.png}
\end{figure}

Figure~\ref{Fig:fig2_mdf.png} shows the MDFs of the chemically defined low-$\alpha$ and high-$\alpha$ populations. The MDFs are normalized to unit area to emphasize differences in their shapes rather than absolute number counts.

The low-$\alpha$ population exhibits a broad MDF with a pronounced peak at mildly subsolar metallicity, around $[\mathrm{Fe}/\mathrm{H}] \sim -0.2$. The distribution extends over a broad metallicity range, with most stars lying approximately between $[\mathrm{Fe}/\mathrm{H}] \sim-0.5$ and $+0.5$, and with a prominent metal-rich tail extending to super-solar values. This large width indicates that the low-$\alpha$ population spans a wide range of chemical enrichment levels. The peak location and overall shape are consistent with results from APOGEE \citep{Hayden2015}, which report a peak around $[\mathrm{Fe}/\mathrm{H}] \sim -0.05$ to $0.0$ in the solar neighbourhood. The presence of a significant metal-rich tail suggests prolonged star formation and sustained chemical enrichment over an extended period in the thin disc \citep{Chiappini1997, Haywood2013}. In addition, radial migration may play an important role in shaping the metal-rich tail of the MDF, as metal-rich stars formed in the inner disc can migrate outwards and populate the solar neighbourhood \citep{Schonrich2009, Minchev2013}.

In contrast, the high-$\alpha$ population displays a noticeably narrower MDF, with a peak located at $[\mathrm{Fe}/\mathrm{H}] \approx -0.5$. The distribution declines rapidly towards higher metallicities, with very few stars reaching near-solar values, and is truncated at the metal-rich end. Its metal-poor tail extends to $[\mathrm{Fe}/\mathrm{H}] \sim -1.25$ \citep{Bensby2003, Reddy2006}. The peak position and narrow width are in good agreement with previous measurements from APOGEE and GALAH \citep{Nidever2014, Buder2019}, indicating that the DESI sample reliably reproduces the well-established chemical structure of the Galactic disc.

Importantly, the two MDFs are well separated, with only a limited overlap near $[\mathrm{Fe}/\mathrm{H}] \sim -0.45$ to $-0.35$. This clear separation demonstrates that the low-$\alpha$ and high-$\alpha$ populations occupy distinct regions in metallicity space and further supports their identification as chemically distinct components of the Galactic disc.

The contrasting shapes of the two MDFs provide quantitative confirmation of their different chemical evolution histories. The broader and more metal-rich MDF of the low-$\alpha$ population is consistent with extended star formation and continuous gas accretion, whereas the narrower and more metal-poor MDF of the high-$\alpha$ population suggests rapid early enrichment followed by a decline in star formation activity \citep{Matteucci2012, Haywood2016}. These differences reinforce the interpretation based on the $[\alpha/\mathrm{Fe}]$--$[\mathrm{Fe}/\mathrm{H}]$ plane that the two populations trace distinct phases of Galactic disc evolution.

\subsection{Vertical metallicity structure and gradients}\label{sec:vertical_gradients}

The vertical distribution of stellar metallicity provides an additional perspective on the chemical structure of the Galactic disc. Having established the bimodal low-$\alpha$ and high-$\alpha$ populations in the previous sections, we next investigated how their metallicity distributions vary with distance from the Galactic mid-plane. In the following analysis, we examine the vertical metallicity gradients of the two populations and explore how their metallicity distributions change with Galactic height.

\paragraph{Vertical metallicity gradients}

\begin{figure}[ht!]
        \centering
        \vspace{-0.2cm}
        \includegraphics[width=\hsize]{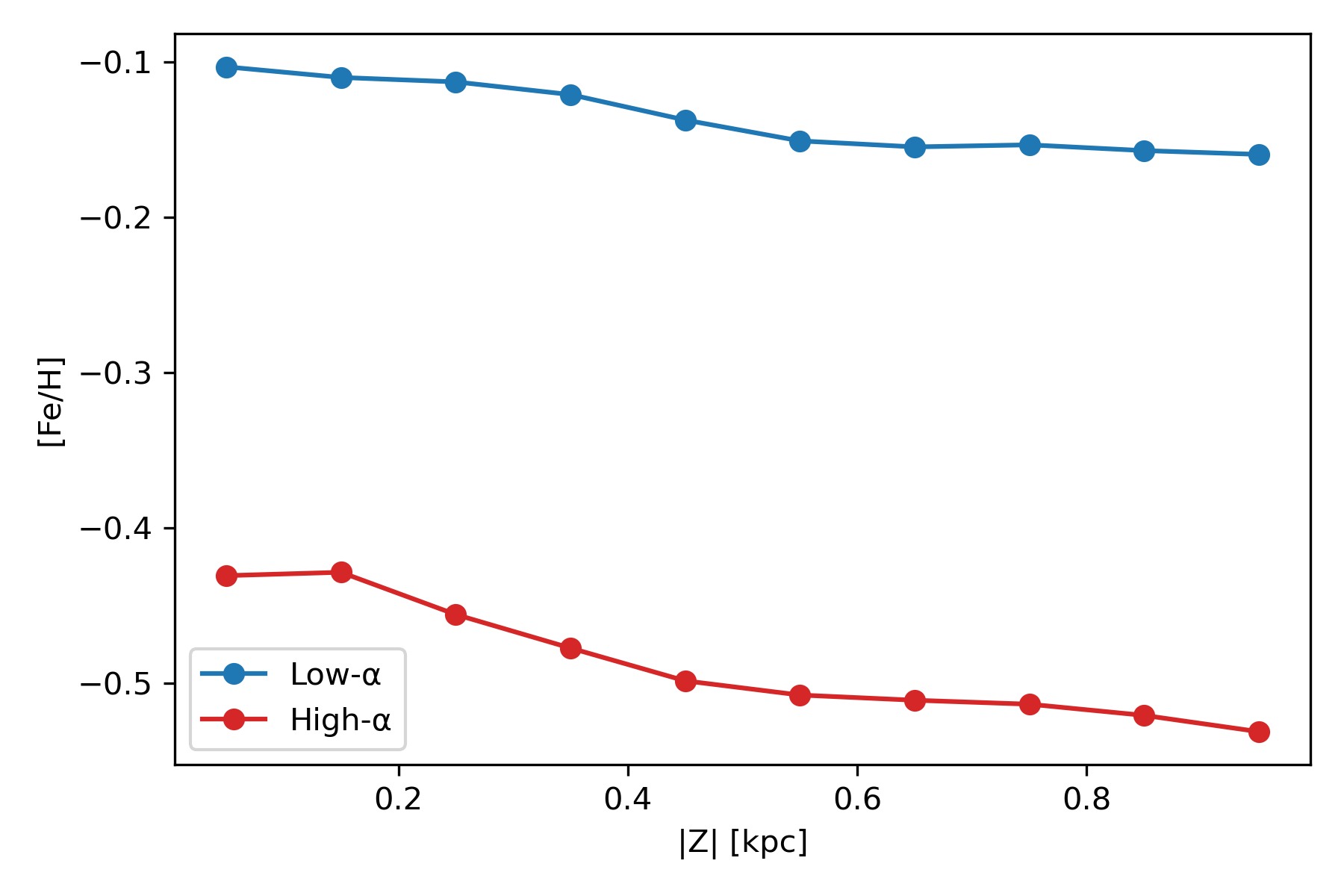}
        \caption{Vertical metallicity gradients as a function of distance from the Galactic mid-plane. The panels show the median $[\mathrm{Fe}/\mathrm{H}]$ in bins of $|Z|$ for the low-$\alpha$ (blue) and high-$\alpha$ (red) populations.}
        \label{Fig:fig3_feh_z.png}
\end{figure}

Figure~\ref{Fig:fig3_feh_z.png} presents the median $[\mathrm{Fe}/\mathrm{H}]$ as a function of $|Z|$ for the low-$\alpha$ and high-$\alpha$ populations. The values are computed in discrete $|Z|$ bins to highlight global trends. The low-$\alpha$ population exhibits a clear negative metallicity gradient,  with the median metallicity decreasing from $[\mathrm{Fe}/\mathrm{H}] \sim -0.10$ near the mid-plane to $\sim -0.16$ at the largest heights probed by our sample. This smooth decline reflects a vertically stratified population and is consistent with previous studies of the thin disc \citep[e.g.][]{Cheng2012, Hayden2015}.

The high-$\alpha$ population also shows a systematic decrease in metallicity with increasing $|Z|$, with the median declining from $[\mathrm{Fe}/\mathrm{H}] \sim -0.43$ to $\sim -0.53$. Quantitatively, the vertical metallicity gradient for the high-$\alpha$ population is steeper than that of the low-$\alpha$ population, consistent with a scenario in which high-$\alpha$ stars formed rapidly in a thicker, dynamically hotter component \citep[e.g.][]{Bensby2014, Hayden2015}.

\begin{figure}[ht!]
        \centering
        \vspace{-0.2cm}
        \includegraphics[width=\hsize]{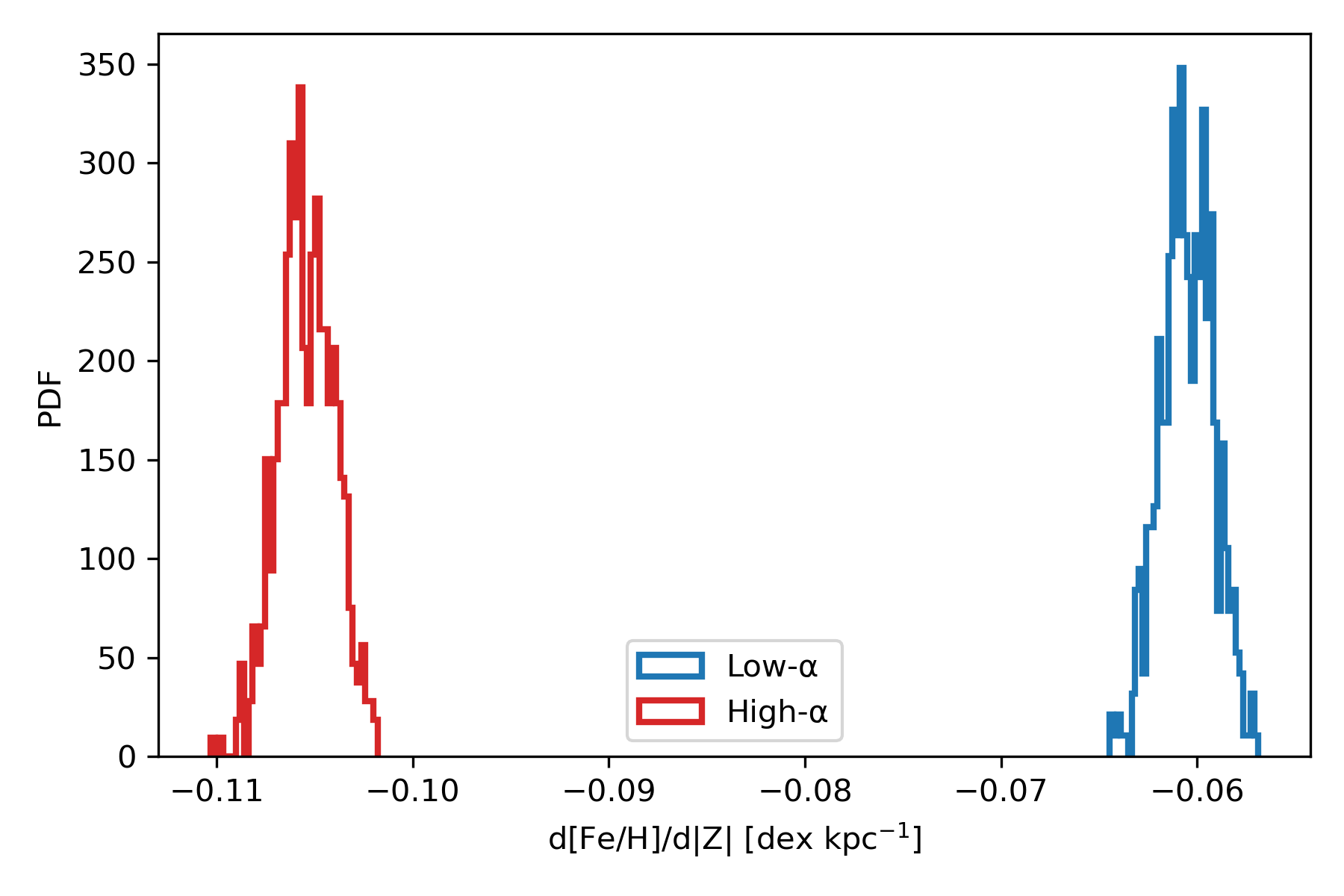}
        \caption{Probability density distributions of the vertical metallicity gradients, $\mathrm{d[Fe/H]}/\mathrm{d}|Z|$, for the chemically defined low-$\alpha$ (blue) and high-$\alpha$ (red) populations in the solar neighbourhood. The distributions are derived from a bootstrap resampling of the stellar sample.}
        \label{Vertical_gradients}
\end{figure}

To evaluate the robustness of the measured gradients, we performed a bootstrap analysis using 1000 resampled realizations. Figure~\ref{Vertical_gradients} shows the resulting distributions of gradient slopes for the low-$\alpha$ and high-$\alpha$ populations. Both distributions are narrow and centred on the best-fitting values derived above, indicating that the inferred gradients are stable against sampling fluctuations. The low-$\alpha$ population exhibits a distribution narrowly peaked around $\sim -0.06$ dex kpc$^{-1}$, whereas the high-$\alpha$ population is centred near $\sim -0.10$ dex kpc$^{-1}$. The systematic offset between the two distributions, corresponding to a difference of approximately 0.04 dex kpc$^{-1}$, demonstrates that the high-$\alpha$ population possesses a significantly steeper vertical metallicity gradient. These results confirm that the negative gradients detected in both populations are statistically robust.

\paragraph{Vertical distributions of the two populations}

\begin{figure}[ht!]
        \centering
        \vspace{-0.2cm}
        \includegraphics[width=\hsize]{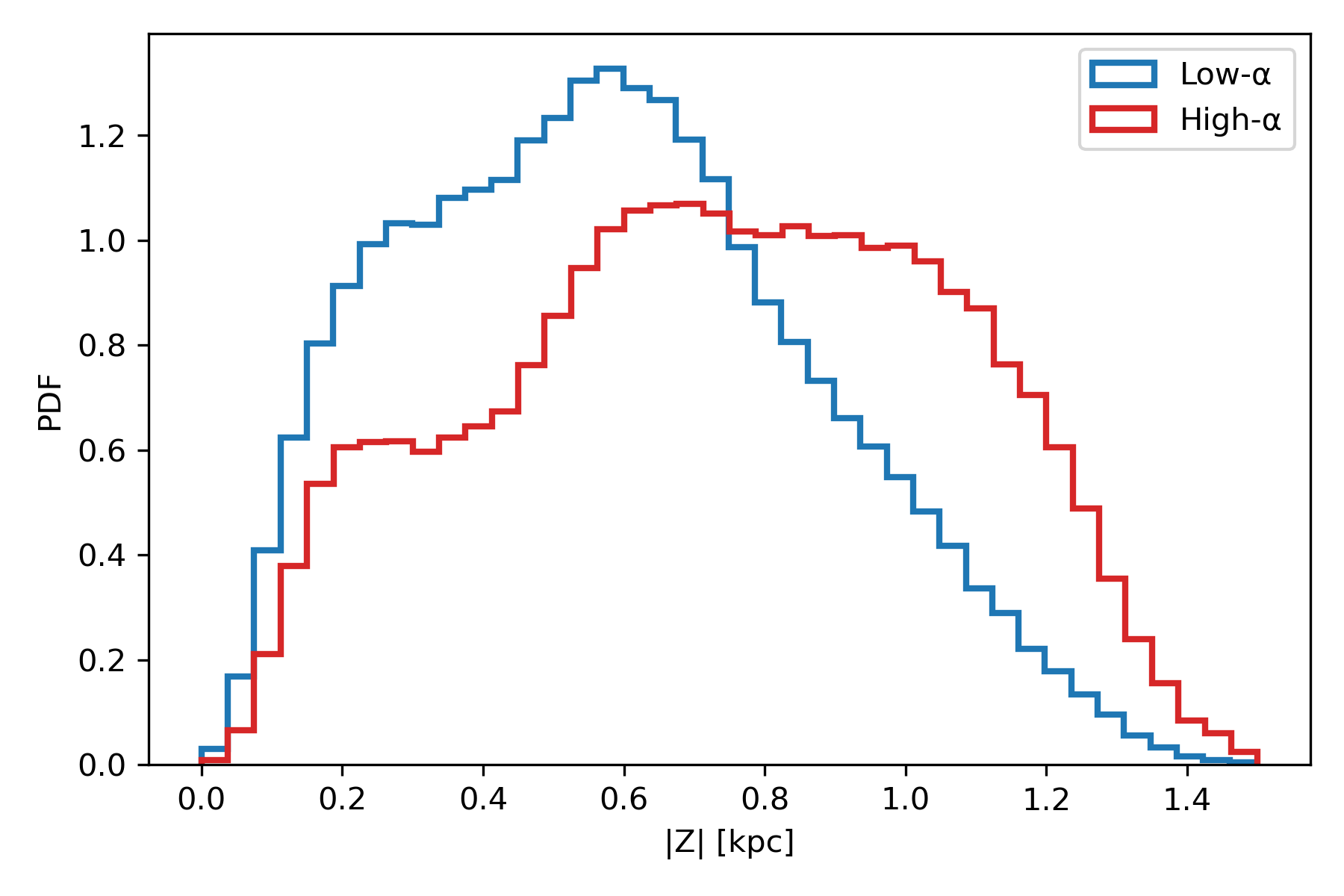}
        \caption{Probability density distributions of the vertical distance, $|Z|,$ for the chemically defined low-$\alpha$ (blue) and high-$\alpha$ (red) populations. The high-$\alpha$ population exhibits a broader and more vertically extended distribution than the low-$\alpha$ population.}
        \label{Fig:fig4_z_dist.png}
\end{figure}

To place the measured vertical metallicity gradients in a broader structural context, we examined the vertical distributions of the chemically defined low-$\alpha$ and high-$\alpha$ populations. Figure~\ref{Fig:fig4_z_dist.png} shows the probability density functions (PDFs) of $|Z|$ for the chemically defined low-$\alpha$ and high-$\alpha$ populations. Although both populations are concentrated within $|Z|<1.5$ kpc, they exhibit systematically different vertical structures. The low-$\alpha$ population is more strongly concentrated towards the Galactic mid-plane, with the highest probability density occurring around $|Z|\sim0.6$ kpc in the present sample. Beyond this peak, the distribution declines rapidly with increasing height. In contrast, the high-$\alpha$ population exhibits a broader vertical distribution, with its peak shifted towards slightly larger $|Z|$ and a more extended tail reaching higher distances from the Galactic plane.

The two distributions overlap substantially over most of the sampled height range. However, the relative contribution of high-$\alpha$ stars increases with increasing $|Z|$, indicating that the high-$\alpha$ population occupies a systematically thicker vertical structure. Together with the vertical metallicity gradients shown in Fig.~\ref{Fig:fig3_feh_z.png}, these results indicate that the low-$\alpha$ and high-$\alpha$ populations differ not only in their chemical properties but also in their spatial distributions. The more vertically extended distribution of the high-$\alpha$ population is consistent with its association with a thicker Galactic disc component, whereas the low-$\alpha$ population is preferentially concentrated towards the Galactic mid-plane \citep[e.g.][]{Bovy2012, Hayden2015}.

Figure~\ref{Fig:fig8_feh_pdf_z_bins.png} provides a more detailed view of the metallicity PDFs in individual $|Z|$ intervals for the low-$\alpha$ and high-$\alpha$ populations. For the low-$\alpha$ population, the distributions exhibit a gradual shift towards lower metallicity with increasing $|Z|$, consistent with the negative vertical metallicity gradient shown in Fig.~\ref{Fig:fig3_feh_z.png}. In addition to the shift in the median metallicity, modest variations in the shape of the distributions are observed across different height intervals, indicating that the observed gradient is associated with both a systematic decrease in metallicity and changes in the metallicity distribution itself.

In contrast, the high-$\alpha$ population exhibits metallicity distributions that remain relatively broad and stable across the sampled $|Z|$ range. Although the median metallicity decreases with increasing height, the overall shape and width of the distributions vary only slightly. This behaviour suggests that the vertical metallicity gradient of the high-$\alpha$ population is primarily reflected in a systematic shift of the metallicity distribution rather than substantial changes in its overall structure. The different behaviours of the two populations are broadly consistent with previous studies of the Galactic thin and thick discs, which have reported distinct vertical chemical structures and metallicity evolution patterns for low-$\alpha$ and high-$\alpha$ stars \citep[e.g.][]{Cheng2012, Hayden2015}.

\makeatletter
\begingroup
\renewcommand*\aa@headcaptionfont{\normalfont}

 \begin{figure*}
        \sidecaption
        \vspace{-0.2cm}
        \includegraphics[width=12cm]{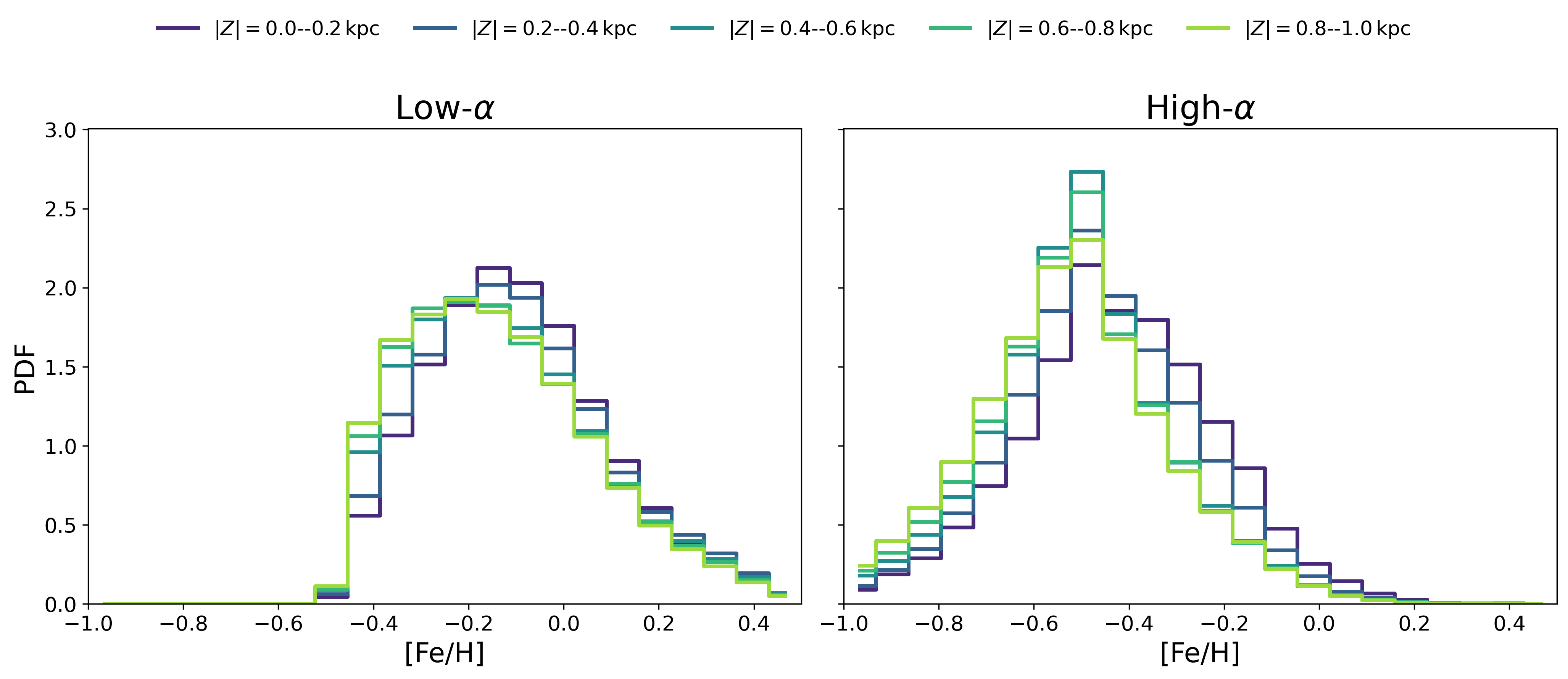}
        \caption{PDFs of $[\mathrm{Fe/H}]$ in different $|Z|$ bins for the low-$\alpha$ (\textit{left}) and high-$\alpha$ (\textit{right}) populations. Each line corresponds to a different vertical height interval above the Galactic mid-plane.}
        \label{Fig:fig8_feh_pdf_z_bins.png}
\end{figure*}

\subsection{Dynamical and radial properties}\label{sec:Dynamical_radial_properties}

To place the vertical metallicity gradients and chemical distributions in a broader Galactic context, we examined the dynamical properties of the chemically distinct low-$\alpha$ and high-$\alpha$ populations. Previous studies have shown that chemically defined disc populations exhibit distinct kinematic properties and occupy different regions of velocity space \citep{Bensby2003, Bovy2012, Haywood2013}. The kinematic behaviour of these populations therefore provides an additional perspective on the connection between chemical abundance patterns and stellar dynamics in the Galactic disc.

Figure~\ref{Fig_fig8_toomre_density_contours_final.png} presents the Toomre diagram of the solar neighbourhood sample. In these diagrams, the azimuthal velocity $V$ is plotted against $\sqrt{U^2 + W^2}$, where $U$ and $W$ are the stellar velocities relative to the local standard of rest in the radial and vertical directions, respectively. The quantity $\sqrt{U^2 + W^2}$ serves as a proxy for the level of dynamical heating, with larger values corresponding to stars on dynamically hotter orbits.

\begin{figure*}
        \sidecaption
        \vspace{-0.2cm}
        \includegraphics[width=12cm]{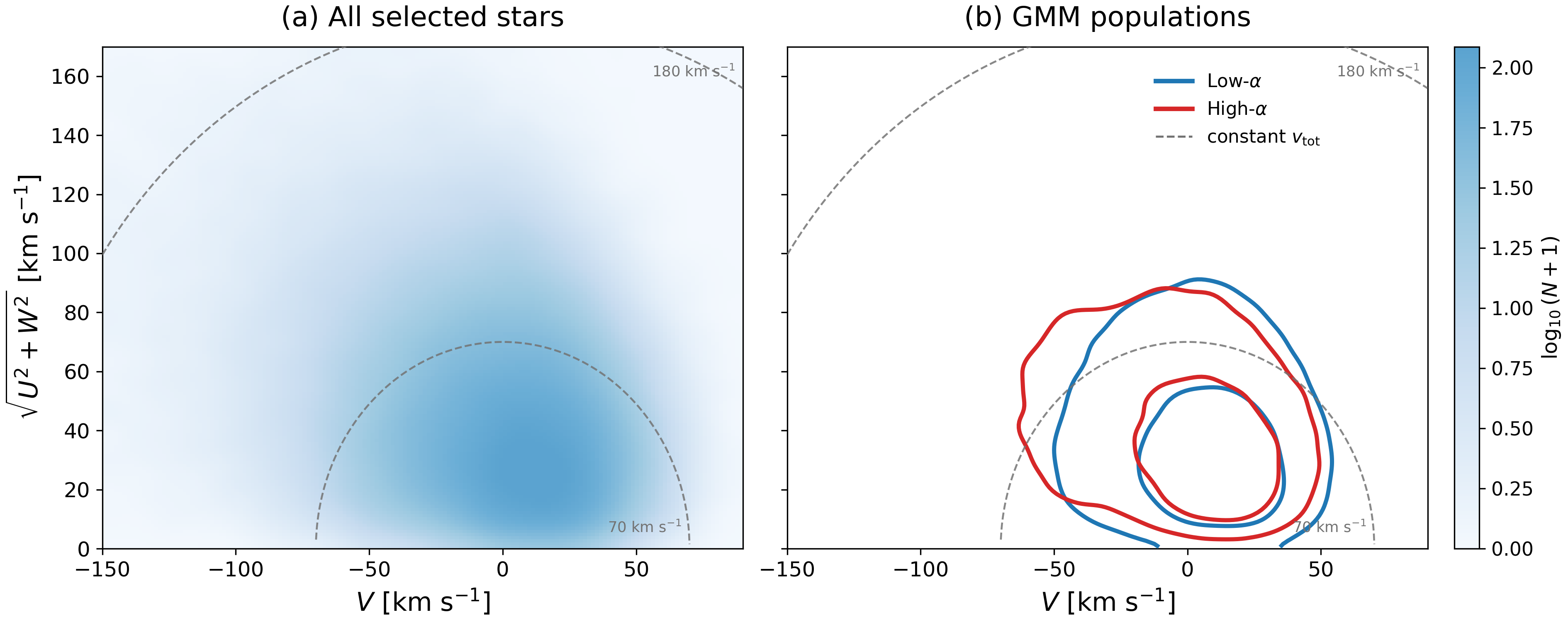}
        \caption{Toomre diagram of the solar neighbourhood sample. \textit{Panel (a)}: Density distribution of all selected stars in the $V$--$\sqrt{U^{2}+W^{2}}$ plane. \textit{Panel (b)}: Density contours of the low-$\alpha$ and high-$\alpha$ populations identified by the GMM decomposition. Dashed curves indicate constant total velocities of 70 and 180 km\,s$^{-1}$.}
        \label{Fig_fig8_toomre_density_contours_final.png}
\end{figure*}

The density distribution of the full sample is strongly concentrated at low values of $\sqrt{U^2+W^2}$ and near $V\approx0~\mathrm{km\,s^{-1}}$, indicating that the solar neighbourhood sample is dominated by dynamically cold disc stars. The stellar density decreases smoothly towards larger values of $\sqrt{U^2+W^2}$ and larger velocity offsets from the local standard of rest.

The kinematic distributions of the low-$\alpha$ and high-$\alpha$ populations exhibit both systematic differences and substantial overlap in the Toomre diagram. The low-$\alpha$ population exhibits a relatively compact distribution concentrated at lower values of $\sqrt{U^{2}+W^{2}}$, whereas the high-$\alpha$ population shows broader contours extending towards larger values of $\sqrt{U^{2}+W^{2}}$ and more negative azimuthal velocities. This contrast demonstrates that the high-$\alpha$ population is dynamically hotter on average and exhibits stronger asymmetric drift than the low-$\alpha$ population, consistent with previous studies of chemically defined Galactic disc populations \citep{Bensby2003, Bovy2012, Nidever2014}. At the same time, the two chemically defined populations overlap substantially in velocity space, suggesting that a considerable fraction of chemically defined high-$\alpha$ stars occupies the same kinematic region as the low-$\alpha$ population.

Quantitatively, within the chemically defined high-$\alpha$ sample, approximately 21.1\% of stars have $V<-50~\mathrm{km\,s^{-1}}$, whereas only 12.4\% have $V<-70~\mathrm{km\,s^{-1}}$. In contrast, about 61.1\% satisfy $V>-50~\mathrm{km\,s^{-1}}$ and $\sqrt{U^2+W^2}<70~\mathrm{km\,s^{-1}}$, indicating that a substantial fraction of the chemically defined high-$\alpha$ population occupies relatively disc-like kinematic regions that overlap with the low-$\alpha$ population despite their distinct $\alpha$-abundance signatures. These thresholds are adopted only to quantify the degree of overlap between chemically defined populations and classical thick-disc-like kinematic regions, rather than to define strict kinematic memberships. These quantitative results indicate that the chemically defined high-$\alpha$ population is dynamically heterogeneous and cannot be described as a purely kinematically hot component. The original stellar distributions used to construct the density contours are presented in Fig.~\ref{FigA1_figA_toomre_scatter_swapped_axes.png}.

Figure~\ref{Fig:Fig_P_high_vs_UW_publication_FINAL.png} further quantifies the connection between chemical membership and stellar kinematics. The background hexagonal density map shows the distribution of stars in the $P_{\mathrm{high}}$--$\sqrt{U^2 + W^2}$ plane, where $P_{\mathrm{high}}$ denotes the GMM membership probability of belonging to the high-$\alpha$ population, while the red points trace the median probability in bins of $\sqrt{U^2 + W^2}$. A statistically significant, moderate monotonic trend ($\rho = 0.21$, $p \ll 10^{-16}$) is observed, indicating that the likelihood of belonging to the high-$\alpha$ population increases with increasing dynamical heating.

A logistic model fitted to the unbinned data provides a smooth representation of this trend. In densely sampled regions, the relation is shown as a solid curve with a shaded $1\sigma$ confidence interval derived from bootstrap resampling. At high $\sqrt{U^2 + W^2}$ values ($\gtrsim 140~\mathrm{km\,s^{-1}}$), where the number of stars is limited, the curve is displayed as dashed to indicate reduced statistical certainty.

These results demonstrate that, although the low-$\alpha$ and high-$\alpha$ populations exhibit systematically different average kinematic properties, they overlap substantially in velocity space. The low-$\alpha$ population is dynamically colder and concentrated near the local standard of rest, whereas the high-$\alpha$ population exhibits larger random motions and stronger asymmetric drift. The smooth increase in $P_{\mathrm{high}}$ with $\sqrt{U^{2}+W^{2}}$, together with the substantial overlap between the two chemically defined populations, indicates that stellar chemistry and present-day stellar kinematics are closely related but are not in one-to-one correspondence.

\begin{figure}[ht!]
        \centering
        \vspace{-0.2cm}
        \includegraphics[width=\hsize]{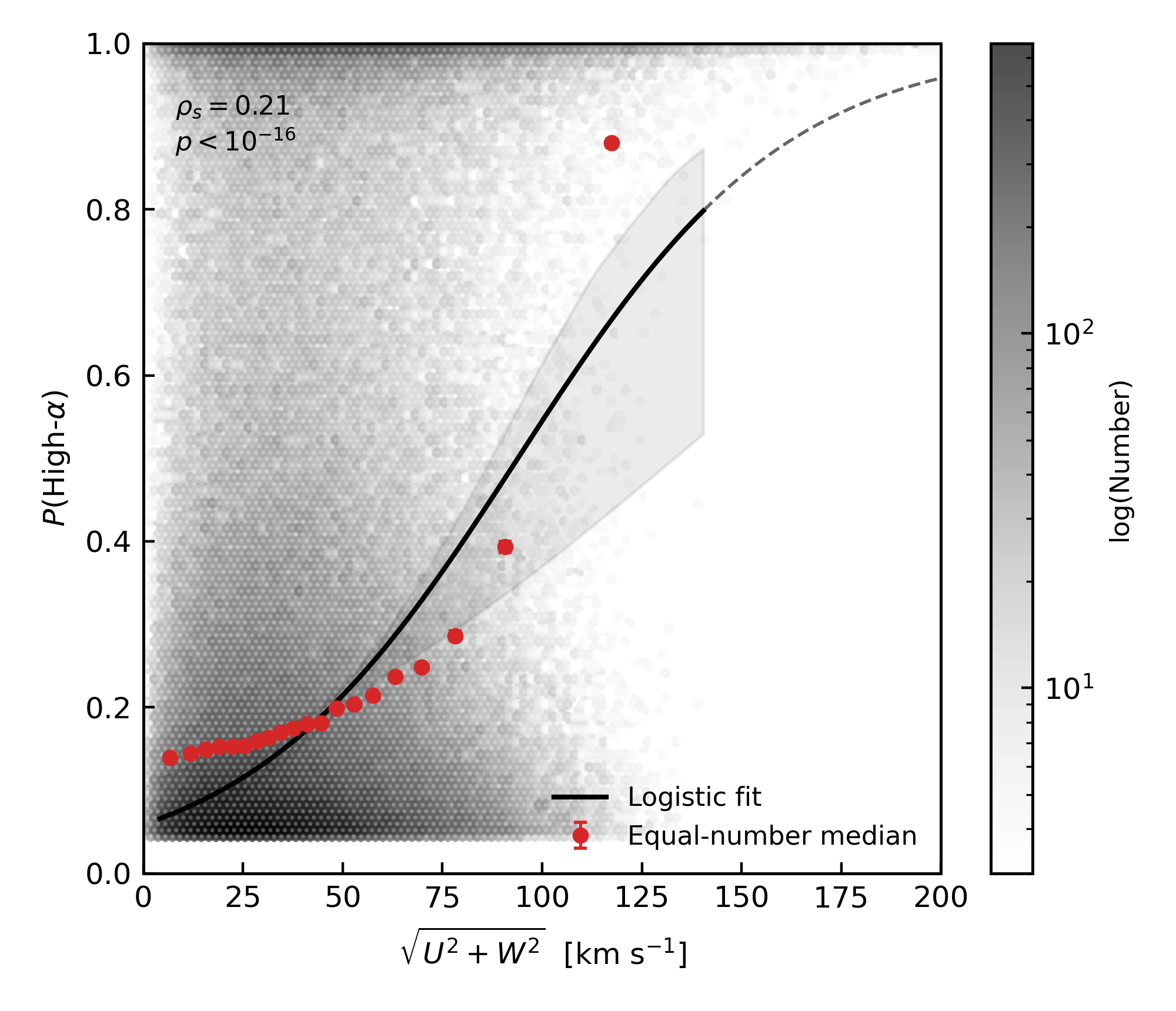}
        \caption{High-$\alpha$ membership probability as a function of the dynamical heating parameter $\sqrt{U^2 + W^2}$. The background grey-scale hexagonal bins show the stellar number density in the $\sqrt{U^2 + W^2}$--$P_{\mathrm{high}}$ plane. Red circles indicate the median $P_{\mathrm{high}}$ values computed in bins of $\sqrt{U^2 + W^2}$. The solid black curve represents the best-fitting logistic model, and the shaded region corresponds to the $1\sigma$ confidence interval derived from bootstrap resampling.}
        \label{Fig:Fig_P_high_vs_UW_publication_FINAL.png}
\end{figure}

Turning to the radial chemical structure, Fig.~\ref{Fig:fig6_r_feh_subplots.png} shows $[\mathrm{Fe/H}]$ as a function of Galactocentric radius $R$ for the two populations, along with the running median in radial bins. The low-$\alpha$ population exhibits a continuous negative radial metallicity gradient, with the median decreasing from near-solar values at $R \sim 7$~kpc to $[\mathrm{Fe/H}] \lesssim -0.2$ at $R \sim 9.6$~kpc. A linear fit to the median yields $\mathrm{d[Fe/H]}/\mathrm{d}R \approx -0.033~\mathrm{dex~kpc^{-1}}$, broadly consistent with previous surveys \citep{Hayden2015, Mackereth2017}. This gradient reflects inside-out disc growth combined with radial migration, which partially redistributes stars across the disc \citep{Schonrich2009, Minchev2013, Frankel2018}.  

 \begin{figure*}
        \sidecaption
        \vspace{-0.2cm}
        \includegraphics[width=12cm]{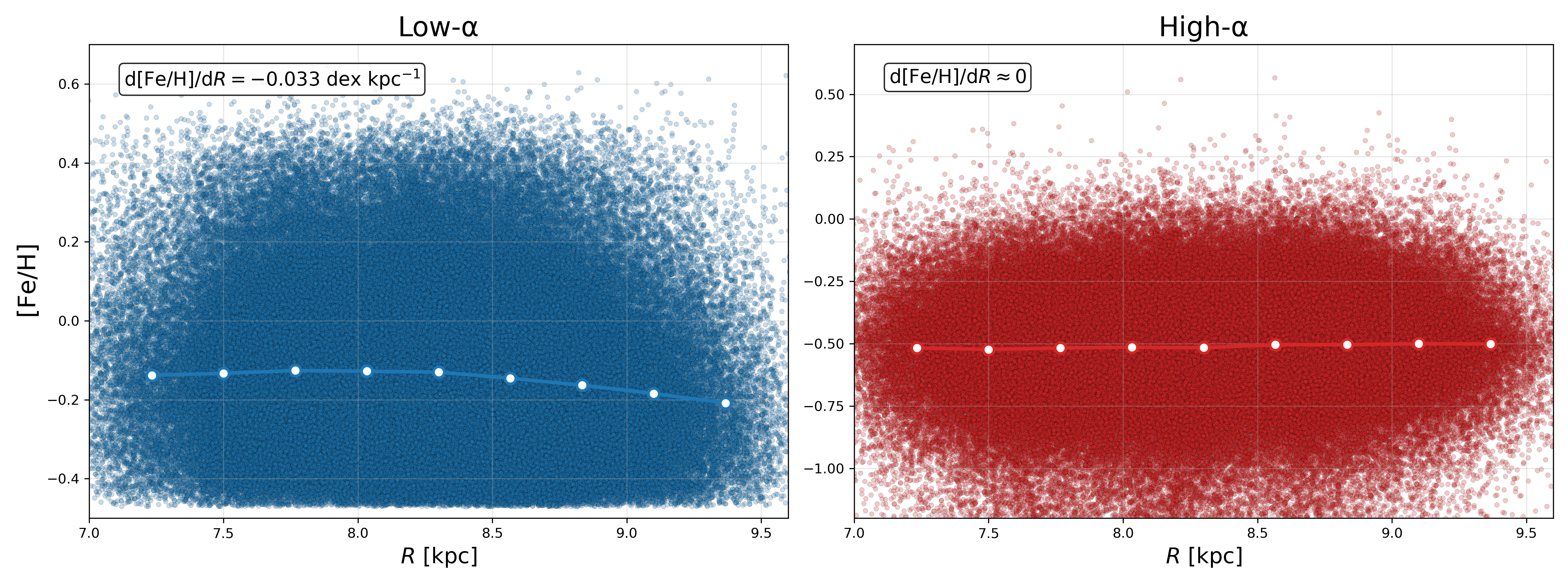}
        \caption{Metallicity as a function of Galactocentric radius for the chemically defined disc populations in the solar neighbourhood. \textit{Left}: Low-$\alpha$ population. \textit{Right}: High-$\alpha$ population. Individual stars are displayed as points, and the white circles connected by solid lines indicate the running median metallicity in radial bins.}
        \label{Fig:fig6_r_feh_subplots.png}
\end{figure*}

In contrast, the high-$\alpha$ population shows an almost flat radial distribution, with median $[\mathrm{Fe/H}] \sim -0.5$ and $\mathrm{d[Fe/H]}/\mathrm{d}R \approx 0$, indicating a chemically well-mixed early disc component. The absence of a significant radial gradient suggests rapid formation and efficient chemical homogenization in the thick disc before substantial radial abundance variations could develop.  

Taken together, the Toomre diagrams, membership probability trends, and radial metallicity distributions reveal a coherent chemical and dynamical picture. The low-$\alpha$ population is dynamically colder and exhibits stronger radial metallicity variations, consistent with a younger disc component formed through prolonged chemical enrichment. In contrast, the high-$\alpha$ population shows, on average, hotter kinematics, weaker radial metallicity variations, and a narrower metallicity distribution, consistent with the properties traditionally associated with the thick disc. However, the substantial fraction of high-$\alpha$ stars occupying relatively cold kinematic regions demonstrates that the chemically defined high-$\alpha$ population is dynamically heterogeneous rather than representing a single kinematically distinct component. These results highlight the coupled but non-equivalent roles of chemical enrichment history and dynamical evolution in shaping the present-day Galactic disc.

\subsection{Summary of the observational results}

The DESI solar neighbourhood sample reveals a clear and robust bimodality in the $[\alpha/\mathrm{Fe}]$--$[\mathrm{Fe}/\mathrm{H}]$ plane, clearly separating two chemically distinct stellar populations, the low-$\alpha$ and high-$\alpha$ sequences. This bimodality persists across the full metallicity range of the sample, confirming the presence of chemically differentiated disc components in the solar neighbourhood \citep[e.g.][]{Hayden2015, Mackereth2019}.

These two populations also exhibit distinctly different metallicity MDFs. The low-$\alpha$ sequence has a broader MDF with a peak near solar metallicity, extending towards both metal-poor and metal-rich extremes, reflecting a range of chemical enrichment levels. In contrast, the high-$\alpha$ population displays a narrower MDF centred at a lower metallicity ($[\mathrm{Fe}/\mathrm{H}] \sim -0.5$), indicative of a population with a relatively narrow metallicity distribution, consistent with a more rapid early enrichment history.

The vertical metallicity structure further differentiates these populations. For the low-$\alpha$ population, the median $[\mathrm{Fe}/\mathrm{H}]$ decreases smoothly with increasing distance from the Galactic mid-plane, and its metallicity distribution shifts towards lower metallicities with increasing Galactic height. In contrast, the high-$\alpha$ population exhibits a steeper vertical metallicity gradient, while the overall shape of its metallicity distribution changes only weakly with Galactic height. Together, these results indicate systematically different vertical metallicity structures, consistent with different chemical enrichment histories for the two chemically defined populations.

Kinematic and radial analyses provide further evidence of systematic differences between the two chemically defined populations. The low-$\alpha$ population is concentrated near the local standard of rest and exhibits a negative radial metallicity gradient, whereas the high-$\alpha$ population is dynamically hotter on average, exhibits a larger asymmetric drift, and displays an approximately flat radial metallicity profile. Nevertheless, the Toomre diagram reveals substantial overlap between the two populations in velocity space, indicating that a considerable fraction of chemically defined high-$\alpha$ stars occupy relatively low-velocity-dispersion regions despite their enhanced $\alpha$ abundances. These observational results indicate that chemically defined populations are not uniquely associated with kinematically distinct Galactic disc components.

Overall, these results demonstrate that the chemically defined low-$\alpha$ and high-$\alpha$ populations differ systematically in their metallicity distributions, vertical distributions, radial and vertical metallicity structures, and kinematic properties, while also exhibiting substantial overlap in velocity space. They provide the empirical foundation for the discussion of the chemo-dynamical evolution of the Galactic disc presented in Sect.~\ref{SEC_Discussion}.

\section{Discussion and conclusions}\label{SEC_Discussion}

The DESI DR1 solar neighbourhood sample reveals a clear chemical bimodality in the Galactic disc, providing a comprehensive view of its chemo-dynamical structure. The bimodal distribution observed in the $[\alpha/\mathrm{Fe}]$--$[\mathrm{Fe}/\mathrm{H}]$ plane confirms the presence of two chemically distinct stellar populations. The low-$\alpha$ population exhibits a broad MDF with a peak around $[\mathrm{Fe}/\mathrm{H}]\sim-0.2$, whereas the high-$\alpha$ population shows a narrower distribution centred near $[\mathrm{Fe}/\mathrm{H}]\sim-0.5$. Similar bimodal chemical sequences have been identified by the APOGEE and GALAH surveys \citep{Hayden2015, Buder2019, Mackereth2019}, demonstrating that the DESI DR1 sample reproduces the fundamental chemical bimodality established by previous large spectroscopic surveys.

The observed chemical bimodality suggests different chemical enrichment histories and star formation timescales for the two populations. The high-$\alpha$ population is generally interpreted as having formed rapidly during the early evolution of the Galactic disc, before Type Ia supernovae contributed substantially to iron enrichment, whereas the low-$\alpha$ population reflects more extended chemical enrichment associated with prolonged star formation \citep{Chiappini1997, Haywood2013}. The observed chemical bimodality therefore provides an important framework for interpreting the differences in spatial distributions and kinematic properties discussed below.

The chemical differences between the two populations are accompanied by systematic differences in their vertical distributions and metallicity structures. Compared with the low-$\alpha$ population, the high-$\alpha$ population exhibits a steeper vertical metallicity gradient and is preferentially distributed at larger distances from the Galactic mid-plane. In contrast, the metallicity distribution function of the low-$\alpha$ population evolves more strongly with Galactic height, suggesting a more complex and spatially diverse chemical evolution history. Together, these properties are consistent with the view that the two chemically defined populations experienced different formation and chemical enrichment pathways.

The radial metallicity profiles provide further evidence that the two chemically defined populations experienced different evolutionary histories. The negative radial metallicity gradient of the low-$\alpha$ population is consistent with the inside-out growth of the Galactic disc, whereas the nearly flat radial metallicity profile of the high-$\alpha$ population suggests weaker radial chemical differentiation compared with the low-$\alpha$ population \citep{Chiappini1997,Hayden2015,Mackereth2019}. Together with the vertical metallicity structures, these radial properties support the view that the chemically defined low-$\alpha$ and high-$\alpha$ populations experienced different formation and chemical evolution pathways.

The kinematic properties, however, reveal a more complex picture than implied by the chemical and metallicity structures alone. Although the high-$\alpha$ population is dynamically hotter on average than the low-$\alpha$ population, a substantial fraction of chemically defined high-$\alpha$ stars occupy the same kinematic region as the low-$\alpha$ population. The degree of this overlap can be quantified from the Toomre diagram. Approximately 21.1\% of the chemically defined high-$\alpha$ stars have $V<-50~\mathrm{km\,s^{-1}}$, whereas only 12.4\% have $V<-70~\mathrm{km\,s^{-1}}$. In contrast, about 61.1\% satisfy $V>-50~\mathrm{km\,s^{-1}}$ and $\sqrt{U^{2}+W^{2}}<70~\mathrm{km\,s^{-1}}$, indicating that a substantial fraction of the chemically defined high-$\alpha$ population occupies relatively low-velocity-dispersion regions that overlap those of the low-$\alpha$ population. These quantitative results demonstrate that chemical and kinematic classifications of Galactic disc populations are not fully equivalent.

These findings indicate that chemically defined high-$\alpha$ stars may occupy markedly different regions of velocity space, suggesting that stellar chemistry and present-day stellar kinematics preserve complementary information about Galactic evolution. This behaviour is consistent with previous studies showing that chemically defined high-$\alpha$ populations cannot be mapped directly onto a single kinematically defined thick-disc component \citep{Bovy2012, Haywood2013}. The coexistence of chemically similar stars with markedly different kinematic properties naturally raises the question of what physical processes are responsible for this diversity. High-$\alpha$ stars with dynamically hotter orbits may be associated with populations that formed during the early turbulent phases of Galactic evolution, whereas those with relatively cold kinematics could have been affected by radial redistribution and other secular processes after their formation. Possible mechanisms include radial migration and long-term dynamical heating, both of which can substantially modify stellar orbital properties while largely preserving stellar chemical abundances \citep{Schonrich2009, Minchev2013, Frankel2018}. Radial migration can redistribute stars over several kiloparsecs, weakening the correspondence between stellar chemistry and present-day orbital properties, whereas secular dynamical heating can progressively increase stellar velocity dispersions without significantly altering stellar chemical abundances. Consequently, similar chemical abundances do not necessarily imply similar present-day kinematic properties.

Although the present data do not allow the relative importance of these evolutionary processes to be quantified, radial migration and dynamical heating provide plausible explanations for the observed kinematic diversity of the chemically defined high-$\alpha$ population. More generally, these results highlight that stellar chemistry primarily preserves information about the chemical conditions of the gas from which stars formed, whereas present-day stellar kinematics encode the cumulative effects of subsequent dynamical evolution. Future studies that combine more precise stellar ages, refined orbital parameters, and larger spectroscopic samples will provide stronger constraints on the relative roles of these processes and help establish how chemical enrichment and dynamical evolution jointly shaped the present-day Galactic disc.

Overall, the DESI DR1 solar neighbourhood sample provides important insights into the chemo-dynamical structure and evolution of the Galactic disc. The combination of chemical abundances, metallicity structures, and stellar kinematics demonstrates that the chemically defined low-$\alpha$ and high-$\alpha$ populations differ systematically in terms of their chemical enrichment histories, metallicity structures, and dynamical properties while also exhibiting substantial overlap in kinematic space. These results demonstrate that chemically defined populations do not correspond uniquely to kinematically defined Galactic disc components but instead preserve complementary information about the chemical enrichment history and subsequent dynamical evolution of the Galactic disc. This study therefore highlights the importance of combining stellar chemistry with stellar kinematics to reconstruct the chemo-dynamical evolution of the Milky Way and provides new observational constraints for future models of Galactic disc formation and evolution.

\section*{Data availability}

This work is based on publicly available data from DESI DR1. The data products used in this study are accessible via the DESI Science Archive:

\begin{itemize}
        \item \url{https://data.desi.lbl.gov/public/dr1/vac/dr1/mws/iron/v1.0/}
        \item \url{https://data.desi.lbl.gov/public/dr1/vac/dr1/mws-specdis/v2.1/}
\end{itemize}

\noindent Additional data related to this work will be made available upon reasonable request.

\begin{acknowledgements}
      This work was supported by the Kunlun Talents Program of Qinghai Province (Grant No.~2025-QLGKLYCZX-026) and by the Qinghai University Research Ability Enhancement Project (Grant No.~2025KTSQ26). This research used data obtained with the Dark Energy Spectroscopic Instrument (DESI). DESI construction and operations are managed by the Lawrence Berkeley National Laboratory. This material is based upon work supported by the U.S. Department of Energy, Office of Science, Office of High-Energy Physics, under Contract No.~DE--AC02--05CH11231, and by the National Energy Research Scientific Computing Center, a DOE Office of Science User Facility under the same contract. Additional support for DESI was provided by the U.S. National Science Foundation (NSF), Division of Astronomical Sciences under Contract No.~AST-0950945 to the NSF’s National Optical-Infrared Astronomy Research Laboratory; the Science and Technology Facilities Council of the United Kingdom; the Gordon and Betty Moore Foundation; the Heising-Simons Foundation; the French Alternative Energies and Atomic Energy Commission (CEA); the National Council of Humanities, Science and Technology of Mexico (CONAHCYT); the Ministry of Science and Innovation of Spain (MICINN), and by the DESI Member Institutions: \url{https://www.desi.lbl.gov/collaborating-institutions}. The DESI collaboration is honored to be permitted to conduct scientific research on I’oligam Du’ag (Kitt Peak), a mountain with particular significance to the Tohono O’odham Nation. Any opinions, findings, and conclusions or recommendations expressed in this material are those of the author(s) and do not necessarily reflect the views of the U.S. National Science Foundation, the U.S. Department of Energy, or any of the listed funding agencies. We sincerely thank the anonymous referee for the careful review of our manuscript and for the constructive and insightful comments and suggestions. The referee's feedback has helped us improve the presentation of the results, clarify the discussion and interpretation, and enhance the overall quality of this manuscript.
\end{acknowledgements}

\begin{appendix}

\nolinenumbers 

\section{Robustness tests}

\subsection{Robustness against sample-selection criteria}\label{Appendix_A1}

To assess the robustness of the chemical bimodality identified by the GMM, we repeated the decomposition using alternative sample-selection criteria in surface gravity, effective temperature, and signal-to-noise ratio. Figure~\ref{Fig_logg_robustness.png} illustrates the results obtained for different surface-gravity cuts, while the quantitative comparisons are summarized in Table~\ref{Tab_gmm_robustness}. In all cases, the low-$\alpha$ and high-$\alpha$ sequences remain clearly visible in the $[\mathrm{Fe}/\mathrm{H}]$--$[\alpha/\mathrm{Fe}]$ plane. The centroid positions and relative population fractions show only minor variations among the different selections. The maximum centroid shift is smaller than 0.013 dex, indicating that the inferred chemical bimodality is insensitive to reasonable changes in the adopted sample-selection criteria. These tests demonstrate that the main conclusions of this work are not driven by the specific choices of $\log g$, $T_{\rm eff}$, or signal-to-noise cuts.

\begin{figure*}[ht!]
        \centering
        \vspace{-0.2cm}
        \includegraphics[width=0.8\textwidth]{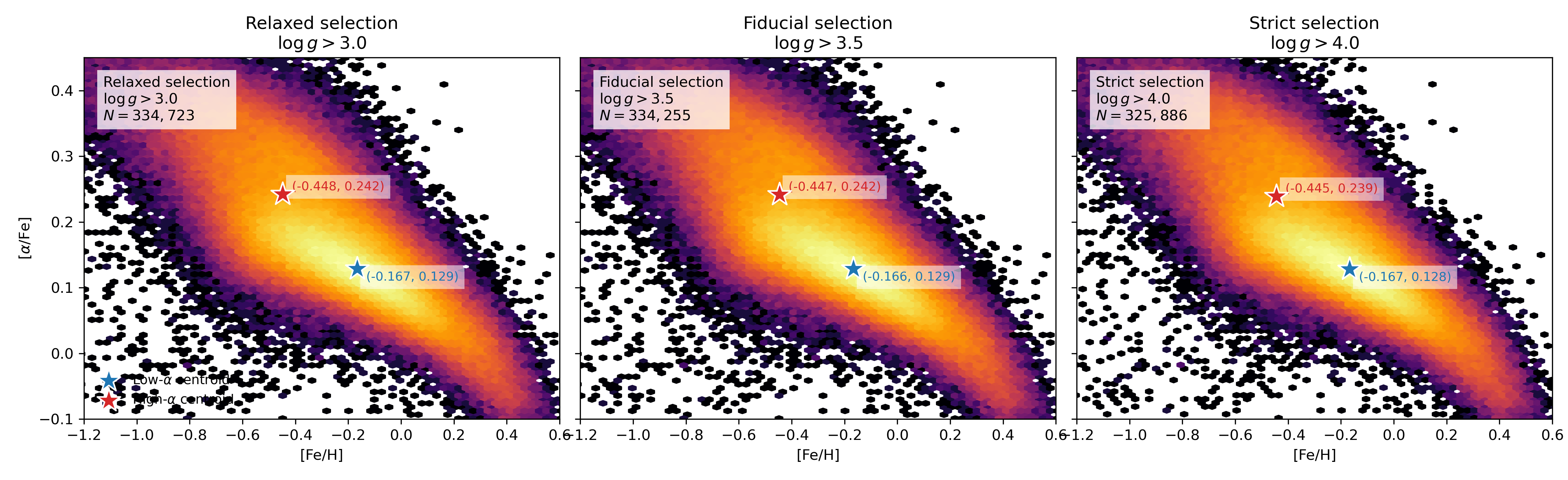}
        \caption{Robustness of the chemical bimodality against variations in the surface-gravity selection criterion. The $[\mathrm{Fe}/\mathrm{H}]$--$[\alpha/\mathrm{Fe}]$ distributions are shown for samples selected with $\log g > 3.0$ (\textit{left}), $\log g > 3.5$ (\textit{middle}), and $\log g > 4.0$ (\textit{right}). Blue and red stars mark the centroids of the low-$\alpha$ and high-$\alpha$ populations identified by the GMM decomposition.}
        \label{Fig_logg_robustness.png}
\end{figure*}

\begin{table*}[!ht]
        \caption{Robustness tests of the GMM decomposition under different sample-selection criteria.}
        \label{Tab_gmm_robustness}
        \centering
        \small
        \begin{tabular}{lcccccc}
                \hline\hline
                Selection &
                $N$ &
                Low-$\alpha$ &
                High-$\alpha$ &
                $f_{\rm low}$ &
                $f_{\rm high}$ &
                $\Delta_{\rm max}$ \\
                \hline
                
                \multicolumn{7}{c}{\textit{Surface-gravity tests}} \\
                \hline
                
                $\log g > 3.0$
                & 334723
                & $(-0.167,0.129)$
                & $(-0.448,0.242)$
                & 59.4
                & 40.6
                & 0.001 \\
                
                Fiducial ($\log g > 3.5$)
                & 334255
                & $(-0.166,0.129)$
                & $(-0.447,0.242)$
                & 59.3
                & 40.7
                & 0.000 \\
                
                $\log g > 4.0$
                & 325886
                & $(-0.167,0.128)$
                & $(-0.445,0.239)$
                & 59.8
                & 40.2
                & 0.003 \\
                
                \hline
                
                \multicolumn{7}{c}{\textit{Effective-temperature tests}}\\
                \hline
                
                $4300 < T_{\rm eff} < 6700$
                & 365490
                & $(-0.165,0.130)$
                & $(-0.434,0.238)$
                & 58.3
                & 41.7
                & 0.013 \\
                
                Fiducial ($4500 < T_{\rm eff} < 6500$)
                & 334255
                & $(-0.166,0.129)$
                & $(-0.447,0.242)$
                & 59.3
                & 40.7
                & 0.000 \\
                
                $4800 < T_{\rm eff} < 6200$
                & 270926
                & $(-0.171,0.128)$
                & $(-0.449,0.236)$
                & 60.9
                & 39.1
                & 0.005 \\
                
                \hline
                
                \multicolumn{7}{c}{\textit{Signal-to-noise tests}} \\
                \hline
                
                $S/N > 20$
                & 334027
                & $(-0.166,0.128)$
                & $(-0.447,0.241)$
                & 59.3
                & 40.7
                & 0.000 \\
                
                Fiducial ($S/N > 10$)
                & 334255
                & $(-0.166,0.129)$
                & $(-0.447,0.242)$
                & 59.3
                & 40.7
                & 0.000 \\
                
                $S/N > 50$
                & 321791
                & $(-0.158,0.126)$
                & $(-0.440,0.238)$
                & 58.2
                & 41.8
                & 0.008 \\
                
                \hline
        \end{tabular}
        
        \tablefoot{
                Low-$\alpha$ and high-$\alpha$ columns list the centroid coordinates in the $[\mathrm{Fe}/\mathrm{H}]$--$[\alpha/\mathrm{Fe}]$ plane. $f_{\rm low}$ and $f_{\rm high}$ denote the population fractions (per cent). $\Delta_{\rm max}$ is the maximum centroid shift relative to the fiducial sample.}
\end{table*}

\subsection{Robustness of the GMM decomposition}\label{Appendix}

To assess the reliability of the GMM decomposition applied to the $[\alpha/\mathrm{Fe}]$--$[\mathrm{Fe}/\mathrm{H}]$ distribution of the solar neighbourhood sample, we performed a series of robustness tests. The purpose of these tests is threefold: (1) to verify that the estimated component centroids are stable under different random initializations of the GMM algorithm, (2) to evaluate whether the adopted two-component model is statistically justified, and (3) to test whether the anti-correlation between $[\alpha/\mathrm{Fe}]$ and $[\mathrm{Fe}/\mathrm{H}]$ is robust against sampling fluctuations.

The analysis was carried out in the two-dimensional chemical abundance space defined by $[\mathrm{Fe}/\mathrm{H}]$ and $[\alpha/\mathrm{Fe}]$. Prior to fitting the mixture model, both variables were standardized using
\begin{equation}
        X_{\mathrm{scaled}} = \frac{X - \mu}{\sigma},
\end{equation}
where $\mu$ and $\sigma$ denote the mean and standard deviation of each variable. This standardization ensures that the dimensions contribute comparably to the clustering and prevents differences in numerical scale from biasing the Gaussian mixture fit.

To examine the stability of the GMM solution, we first performed a centroid stability test. A two-component GMM was repeatedly fitted to the standardized chemical space using different random initializations. Each run was initialized with a unique random seed, and the procedure was repeated 120 times. The resulting centroids of the two Gaussian components were transformed back to the original $[\alpha/\mathrm{Fe}]$--$[\mathrm{Fe}/\mathrm{H}]$ space for comparison. As shown in Fig.~\ref{FigA1_GMM_Robustness.png}a, the centroids cluster tightly, demonstrating that the GMM solution is insensitive to initialization and that the locations of the two chemical components are well constrained by the data.

We next investigated the appropriate number of mixture components by comparing models of increasing complexity. GMMs with one to five components were fitted, and the BIC was calculated for each model. The normalized difference,
\begin{equation}
        \Delta \mathrm{BIC} = \mathrm{BIC}_k - \mathrm{BIC}_{\min},
\end{equation}
was used to facilitate comparison, where $\mathrm{BIC}_k$ is the BIC for a model with $k$ components and $\mathrm{BIC}_{\min}$ is the minimum BIC among all tested models. Figure~\ref{FigA1_GMM_Robustness.png}b shows $\Delta \mathrm{BIC}/N$ as a function of component number. The minimum BIC occurs for the two-component model, confirming that it provides the optimal balance between goodness-of-fit and model complexity. Adding more components yields only marginal improvement, indicating that the dominant structure of the chemical abundance distribution is adequately captured by two Gaussians.

Finally, we evaluated the robustness of the anti-correlation between $[\alpha/\mathrm{Fe}]$ and $[\mathrm{Fe}/\mathrm{H}]$ using bootstrap resampling. We generated 800 bootstrap realizations of the stellar sample by randomly resampling the stars with replacement. For each realization, the Spearman rank correlation coefficient $\rho$ between $[\mathrm{Fe}/\mathrm{H}]$ and $[\alpha/\mathrm{Fe}]$ was calculated. As shown in Fig.~\ref{FigA1_GMM_Robustness.png}c, the bootstrap distribution of the Spearman correlation coefficient is narrowly concentrated around a median value of $\rho \approx -0.796$. This demonstrates that the anti-correlation is strong and stable and is not driven by sampling fluctuations.

\begin{figure*}[ht!]
        \centering
        \vspace{-0.2cm}
        \includegraphics[width=0.8\textwidth]{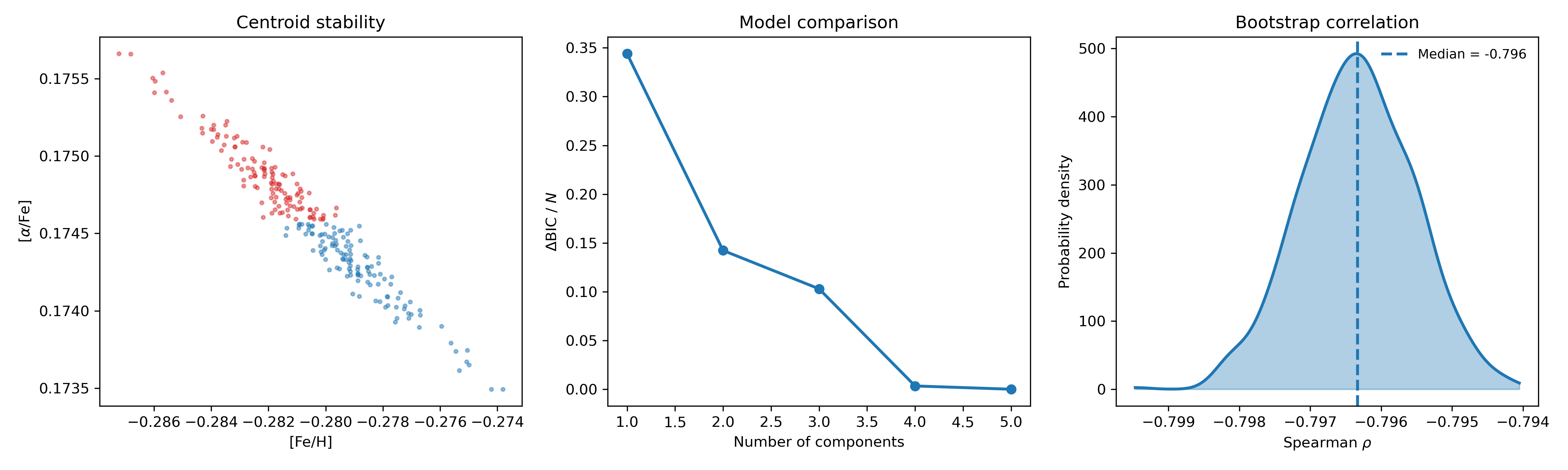}
        \caption{Robustness tests of the GMM decomposition in the [$\alpha$/Fe]--[Fe/H] plane. \textit{Left}: Centroid stability test showing the distribution of low-$\alpha$ (blue) and high-$\alpha$ (red) centroids from 120 independent random initializations. \textit{Middle}: Model comparison using the BIC for GMMs with 1–5 components; the two-component model minimizes the BIC. \textit{Right}: Kernel density estimate of the bootstrap Spearman correlation coefficient ($\rho$) between [$\alpha$/Fe] and [Fe/H] from 800 bootstrap realizations.}
        \label{FigA1_GMM_Robustness.png}
\end{figure*}

\section{Toomre diagrams of the low-$\alpha$ and high-$\alpha$ populations}\label{Appendix_B}

For completeness, Fig.~\ref{FigA1_figA_toomre_scatter_swapped_axes.png} presents the original Toomre diagrams of the low-$\alpha$ and high-$\alpha$ populations shown as individual stars. These distributions correspond to the stellar samples used to construct the density contours displayed in Fig.~\ref{Fig_fig8_toomre_density_contours_final.png}. The same velocity definitions adopted in the main text are used here, where $V$ denotes the azimuthal velocity relative to the local standard of rest and $\sqrt{U^2+W^2}$ characterizes the velocity component perpendicular to the direction of Galactic rotation. Dashed curves indicate constant total velocities of 70 and 180 km s$^{-1}$.

The original stellar distributions confirm the trends identified in the density and contour representations. The low-$\alpha$ population is concentrated towards lower values of $\sqrt{U^2+W^2}$, whereas the high-$\alpha$ population occupies a broader region of velocity space and extends to larger values of $\sqrt{U^2+W^2}$. Nevertheless, the two populations exhibit substantial overlap, consistent with the continuous transition observed in the main-text analysis.

 \begin{figure*}
        \sidecaption
        \vspace{-0.2cm}
        \includegraphics[width=12cm]{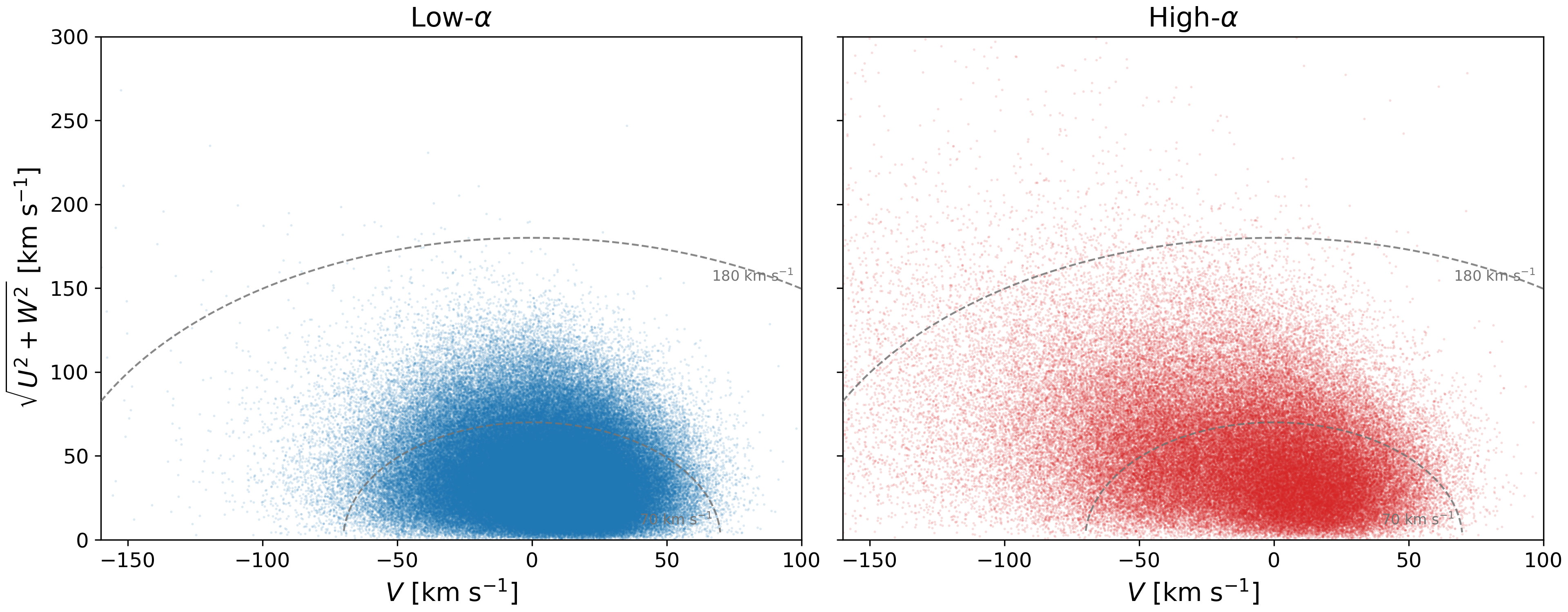}
        \caption{Toomre diagrams of the low-$\alpha$ (\textit{left}) and high-$\alpha$ (\textit{right}) populations shown as individual stars in the $V$--$\sqrt{U^{2}+W^{2}}$ plane. Dashed curves indicate constant total velocities of 70 and 180 km\,s$^{-1}$.}
        \label{FigA1_figA_toomre_scatter_swapped_axes.png}
\end{figure*}

\FloatBarrier 
\clearpage

\end{appendix}
\end{document}